\documentclass{article} 
\usepackage{iclr2027_conference,times}
\usepackage{upgreek}

\usepackage{amsmath,amsfonts,bm}

\def\eqref#1{equation~\ref{#1}}

\def\1{\bm{1}}

\def\vb{{\bm{b}}}
\def\vc{{\bm{c}}}

\def\ve{{\bm{e}}}
\def\vf{{\bm{f}}}

\def\vh{{\bm{h}}}

\def\vq{{\bm{q}}}
\def\vr{{\bm{r}}}

\def\vu{{\bm{u}}}
\def\vv{{\bm{v}}}

\def\vx{{\bm{x}}}
\def\vy{{\bm{y}}}
\def\vz{{\bm{z}}}

\def\mA{{\bm{A}}}
\def\mB{{\bm{B}}}

\def\mD{{\bm{D}}}

\def\mI{{\bm{I}}}
\def\mJ{{\bm{J}}}

\def\mM{{\bm{M}}}

\def\mQ{{\bm{Q}}}
\def\mR{{\bm{R}}}
\def\mS{{\bm{S}}}
\def\mT{{\bm{T}}}
\def\mU{{\bm{U}}}

\def\mW{{\bm{W}}}

\DeclareMathAlphabet{\mathsfit}{\encodingdefault}{\sfdefault}{m}{sl}
\SetMathAlphabet{\mathsfit}{bold}{\encodingdefault}{\sfdefault}{bx}{n}

\DeclareMathOperator{\Tr}{Tr}

\usepackage[dvipsnames]{xcolor}
\usepackage{hyperref}
\hypersetup{
  colorlinks=true,
allcolors=ForestGreen!80!black}%
\usepackage{url}

\usepackage{graphicx}
\usepackage{amsmath}
\usepackage{amsfonts}
\usepackage{amssymb}
\usepackage{wrapfig}

\usepackage{chngcntr}
\usepackage{caption} 
\usepackage{titlesec}

\titlespacing*{\section}{0pt}{4pt plus 0pt minus 2pt}{4pt plus 0pt minus 1pt}
\titlespacing*{\subsection}{0pt}{1.5pt plus 0pt minus 2pt}{1.5pt plus 0pt minus 1pt}

\title{Efficiency and robustness partition the solution space for memory in recurrent networks}

\author{William Qian$^{1,2}$, Cengiz Pehlevan$^{2,3,4}$ \\
\textsuperscript{1}Biophysics Graduate Program\\
\textsuperscript{2}Kempner Institute for the Study of Natural and Artificial Intelligence  \\
\textsuperscript{3}John A. Paulson School of Engineering and Applied Sciences \\
\textsuperscript{4}Center for Brain Science \\
Harvard University, Cambridge MA 02138 \\
\texttt{williamqian@g.harvard.edu, cpehlevan@seas.harvard.edu} \\
}

\iclrfinalcopy 
\begin{document}

\maketitle

\begin{abstract}
In computational neuroscience, task-trained recurrent neural networks (RNNs) are commonly used as a testbed for exploring the space of recurrent circuit solutions compatible with a particular function. However, these networks are subject to inductive biases that may be misaligned with those of biological circuits, which operate under various efficiency and robustness constraints. Here, using a minimal stimulus recall task, we systematically characterize how the solution space of recurrent neural networks is shaped by the desiderata of weight efficiency, activity efficiency, and robustness to noise. We show that weight efficient solutions exhibit a low-rank bias that aligns with the inductive biases of task-trained networks and favors dynamical motifs involving slow, persistent dynamics, whereas activity efficiency encourages high-rank solutions that rely on transient amplification. Moreover, we find that activity efficient solutions exhibit a strong dissociation between high-variance modes of activity and those that causally influence network output. Unlike in static feedforward settings, we demonstrate that noise robustness and activity efficiency are generally in opposition, rather than synergistic. Using these findings, we propose a normative account of misaligned coding phenomena recently identified in premotor cortex. 

\end{abstract}

\section{Introduction}
\setlength{\abovedisplayskip}{3pt}
\setlength{\belowdisplayskip}{3pt}
A central goal of neuroscience is to understand the representations and dynamical motifs that underlie neural computation, and why they might arise. A large body of work seeks to use task-optimized artificial neural networks (ANNs) to address this question \citep{manteContextdependentComputationRecurrent2013,yaminsPerformanceoptimizedHierarchicalModels2014,rajanRecurrentNetworkModels2016,sorscherUnifiedTheoryOrigin2019,richardsDeepLearningFramework2019,schrimpfBrainScoreWhichArtificial2020}. In many cases, task-trained ANNs have been argued to recapitulate key features of computations in biological brains, despite obvious mismatches in their computational substrates and learning rules. This perspective, while enjoying remarkable success in recent years, still leaves unanswered the question of why a particular solution has been chosen by biology when many
task-compatible solutions exist.

Relatedly, a growing body of work has sought to characterize the solution space of RNNs trained on neuroscience-style tasks, with variation induced by random seeds, architectural variations, learning rules, feature learning strength, or explicit diversity-seeking regularizers \citep{maheswaranathanUniversalityIndividualityNeural2019,turnerChartingNavigatingSpace2021,kurtkayaDynamicalPhasesShortterm2025,huangMeasuringControllingSolution2025,qianDiscoveringAlternativeSolutions2025,paganIndividualVariabilityNeural2025,bordelonDynamicsLearningIntegrate2026, clarkStructureDisorderDynamics2026,ostrowMetricComparingComplex2026}. Among the vast sea of solutions, what organizing principles select the solutions actually implemented in biological circuits? 
Normative theories such as efficient coding propose that neural circuits select metabolically efficient solutions, an insight that is typically operationalized through constraints on activation sparsity, firing rates, or synaptic weight norms \citep{attneaveInformationalAspectsVisual1954}. This perspective has most famously been used to explain sensory representations in visual cortex \citep{olshausenNaturalImageStatistics1996,olshausenSparseCodingOvercomplete1997}. More recently, similar ideas have been extended beyond sensory coding to neural computations supporting cognition and behavior \citep{motiwalaEfficientCodingCognitive2022,schuttRewardPredictionError2024,stroudEffectsNoiseMetabolic2025,dorrellEfficientComputingTheory2026}. While the effects of efficiency and robustness constraints have been extensively studied in feedforward networks \citep{atickTheoryEarlyVisual1990,bishopTrainingNoiseEquivalent1995,karklinEfficientCodingNatural2011,whittingtonDisentanglementBiologicalConstraints2022,braunNotAllSolutions2025}, a precise understanding of how these desiderata shape the solution space of recurrent circuits is still lacking. 

Standard procedures for training RNNs on neuroscience-style tasks already favor one particular notion of efficiency--small weight norm--whether implicitly through the inductive biases of gradient-based optimization or through explicit regularization. In particular, task-trained RNNs often learn minimal solutions with approximately low-rank structure in their recurrent weights, with rank roughly proportional to task dimensionality \citep{schuesslerInterplayRandomnessStructure2020,bordelonDynamicsLearningIntegrate2026, clarkStructureDisorderDynamics2026}. Often, this low-dimensional description of neural dynamics seems apt, especially when the effective dimensionality of neural activity observed in experimental recordings is small. Yet, this perspective can fail to capture important structure when modes of activity that are highly functionally relevant contribute only a small fraction of the explained variance, and are instead discarded as task-irrelevant noise. Indeed, recent work has shown that dynamics in anterior lateral motor cortex (ALM) during a memory-guided motor task are well-described by task coding directions, which explain a large fraction of the variance in activity, and low-variance residual directions, which ultimately exert outsized behavioral influence \citep{daieFeedforwardAmplificationRecurrent2023,pereira-obilinovicNeuralDynamicsOutside2026}. 
Why might this ``misaligned" coding be favorable?

Motivated by these observations, we study how  weight efficiency, activity efficiency, and noise robustness partition the space of recurrent solutions to memory functions, using these desiderata as interpretable axes along which to evaluate solutions.  
Under an analytically tractable model of stimulus recall, we show that weight efficiency pushes networks towards a minimal, rank-1 solution where stimulus identity, along with input, readout, and recurrent weights align to a common direction, coinciding with the simplicity bias of standard task-trained networks. In contrast, all of these components become misaligned in activity efficient solutions, which instead use high rank recurrent weights to transiently amplify the stimulus in an encoding distributed across many modes of activity. Further, we show that noise robustness constraints set fundamental limits on how activity efficient a solution can be. We demonstrate that these findings extend qualitatively to more complex settings in nonlinear RNNs. Finally, by varying solutions along these axes, we show that pairing noise robustness with an intermediate level of activity efficiency best recapitulates the misaligned coding phenomena described above in a model of anterior lateral motor cortex.

\section{The solution space of stimulus recall}

\begin{figure}
    \centering
    \includegraphics[width=\linewidth]{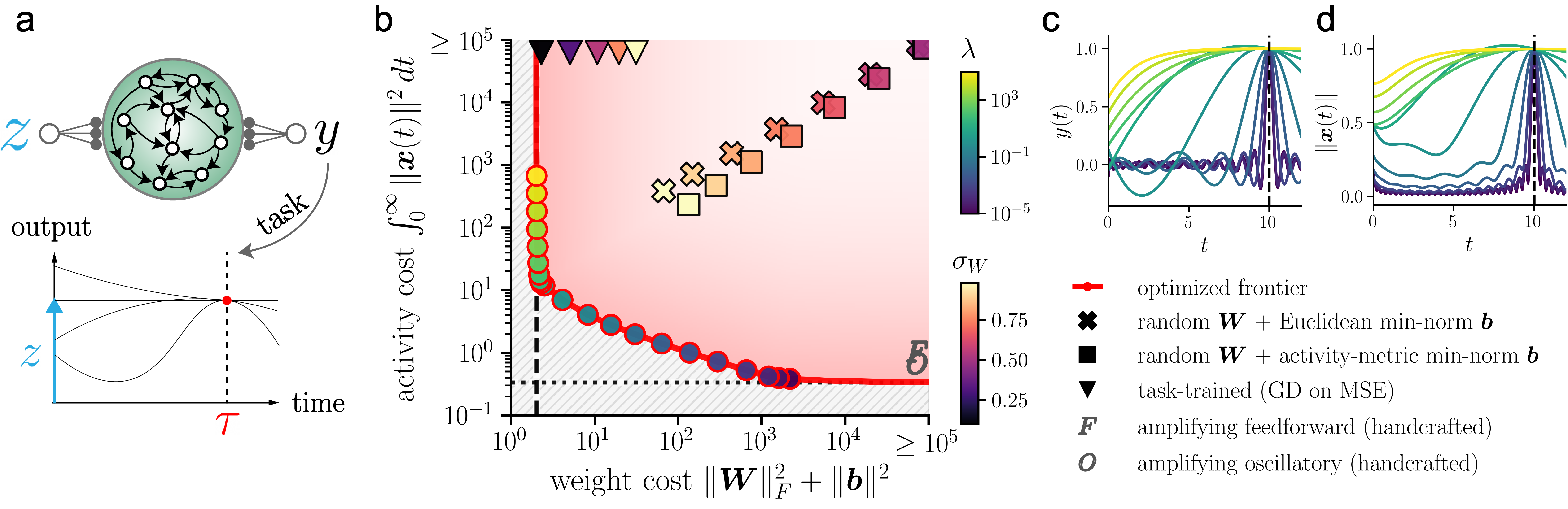}
    \caption{\textbf{Characterizing the space of recurrent solutions to stimulus recall by their efficiency.} \textbf{a.} Schematic for the stimulus recall task, showing multiple task-compatible output traces. \textbf{b.} Weight and activity costs across solution families, including networks with unstructured recurrent weights, task-trained networks, cost-optimized networks, and handcrafted networks. Where relevant, $\sigma_W$ indicates the initialization scale of the recurrent weights. The red boundary indicates the optimized frontier obtained over $\lambda$. Dashed lines indicate lower bounds on the costs. \textbf{c, d.} Readout and activity norm traces across cost-optimized networks at different $\lambda$, shown for unit input $z=1$. }
    \label{fig:1}
\end{figure}

We begin by studying the solution space of a simple stimulus recall task (Fig. \ref{fig:1}a), which we have formulated to be deliberately minimal so as to aid both analytical tractability and conceptual clarity. This setting isolates the maintenance of task-relevant information across a delay, a basic computational motif underlying memory-guided behavior. 
Consider a scalar input-output mapping governed by a linear RNN:
\begin{align}
    \dot{\vx} &= -\vx + \mW\vx + \vb u(t), \\
    y(t) &= \vc ^\top \vx(t),
\end{align}
where $\mW \in \mathbb{R}^{N\times N}$, $\vb \in \mathbb{R}^{N}$, and $\vc \in \mathbb{R}^N$ represent the recurrent, input, and readout weights, respectively. In each trial, the network receives a pulse input $u(t) = z\delta(t)$, where $z \sim P(z)$ represents an observed scalar stimulus, and the network is simply required to reproduce the observed stimulus $z$ after a delay of time $\tau$. Crucially, we do not explicitly constrain the readout $y(t)$ at any time other than the readout time $\tau$. Solving this task then amounts to the condition $y(\tau) = z$ for any stimulus $z$, or equivalently, requiring that $g(\tau) =\vc^\top e^{(-\mI+\mW)\tau}\vb = 1$, where $g(\tau)$ is the input-to-output gain at time $\tau$.

A natural way to find a solution for this task is to minimize the mean squared error $\mathbb{E}_z[(y(\tau)-z)^2]$ via gradient descent over the parameters $\bm{\Theta}=\{ \mW, \vb, \vc \}$. More generally however, the solution space for this task is vastly degenerate; minimizing MSE via gradient descent only imposes a particular tiebreaker for this degeneracy. It is well known that gradient descent has an implicit bias towards simple solutions, and in particular solutions with small parameter norm \citep{neyshaburSearchRealInductive2014,gunasekarCharacterizingImplicitBias2018}. Indeed, although the standard minimum-norm result for linear regression does not apply because the map from $\bm{\Theta}$ to $y(\tau) = g(\tau) z$ is nonlinear and the resulting MSE objective is nonconvex, we show in App. \ref{sec:gd} that gradient descent from small initialization recovers an aligned, rank-1 solution similar to the minimum norm solution (under similar assumptions as in \citep{bordelonDynamicsLearningIntegrate2026}, detailed in \ref{sec:gd}). 

However, biological circuits operate under metabolic and energetic constraints that may not necessarily align with the simplicity-seeking bias of gradient descent. In particular, solutions that require large weights or excess network activity are both potentially undesirable. Thus, we consider the constrained optimization problem
\begin{equation}\label{eqn:problem}
    \min_{\mW,\vb} \quad \mathbb{E}_z\left[\int_0^\infty \|\vx(t)\|^2 dt\right] + \lambda \left(\|\mW\|_F^2 + \|\vb\|^2\right) \qquad \text{s.t. } g(\tau) = 1,
\end{equation}
where $\lambda$ is a parameter governing the relative importance of the activity and weight costs. 
The activity cost is defined with an infinite upper limit to model the total activity over a long horizon, and converges to a finite value so long as $\mW$ is asymptotically stable (i.e., eigenvalue spectrum has maximum real part less than unity). We fix a choice of readout vector $\vc$ with $\|\vc\| = 1$, which can be justified for two reasons: by rotational symmetry of linear RNNs, we can always fix the direction of $\vc$ without loss of generality. We choose to also fix the readout norm because in the highly activity-constrained regime $\lambda \rightarrow 0$, the scaling symmetry $\vc \rightarrow \alpha \vc$, $\vb \rightarrow \alpha^{-1} \vb$ would allow for arbitrarily small activity by making the readout vector large and the input vector tiny, thereby making the optimization problem ill-posed. Thus, we require activity efficiency independent of output scale. 

This constrained optimization problem can be reduced to the unconstrained optimization $\min_{\mW} \mathcal{L}(\mW)$, with an objective dependent solely on the recurrent weights $\mW$ (App. \ref{sec:reduced_problem}):
\begin{equation}\label{eqn:reduced}
    \mathcal{L}(\mW) = \lambda \|\mW\|^2_F + \frac{1}{\vq(\tau)^\top (\mR+\lambda \mI)^{-1}\vq(\tau)},
\end{equation}
where we define the effective dynamics matrix $\mA = -\mI+\mW$, the Gramian matrix $\mR = \int_0^{\infty} e^{\mA^\top t}e^{\mA t} dt$, the adjoint state $\vq(\tau) = e^{\mA^\top \tau}\vc$, and without loss of generality have taken the stimulus variance as $\sigma^2_z=1$.  Here, we have integrated out the optimal input weights $\vb^\star \propto (\mR + \lambda \mI)^{-1}\vq(\tau)$, normalized to satisfy $g(\tau)=1$. 
For fixed recurrent weights, the second term in Eq. \ref{eqn:reduced} can be interpreted as the minimum combined input weight and activity cost necessary to solve the task.
\begin{figure}[t]
    \centering
    \includegraphics[width=0.85\linewidth]{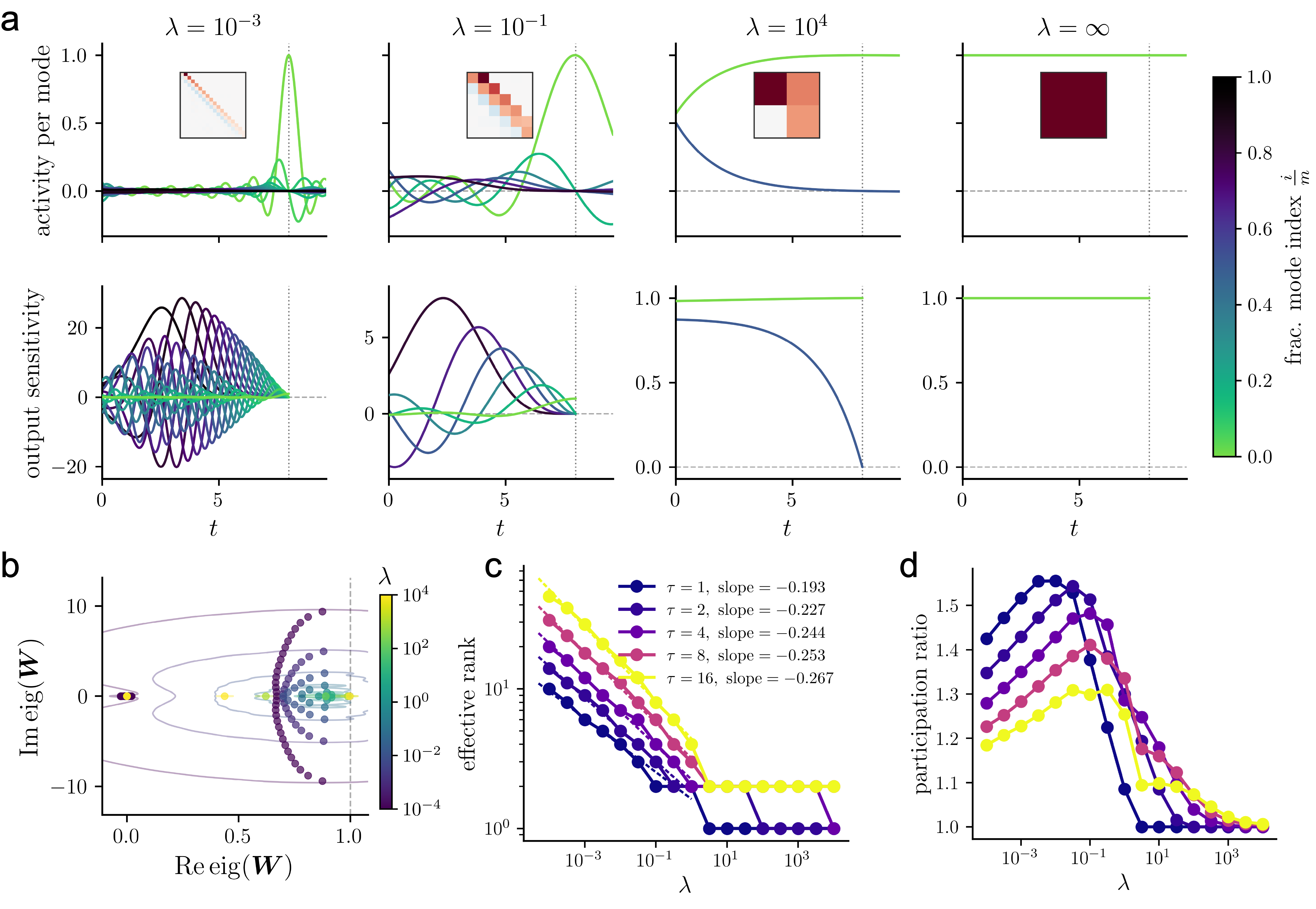}
    \caption{\textbf{Activity efficiency promotes high-rank, transiently amplifying solutions characterized by a dissociation between explained variance and causal influence.} \textbf{a.} Top: Per-mode activity of optimized networks at different $\lambda$, with modes defined by the Arnoldi basis. Traces are colored by their zero-indexed fractional mode index, so that light green corresponds to the first mode, the readout direction. Insets show the recurrent weights in this basis. Dotted lines indicate the readout time. Bottom: Sensitivity of the output at the readout time with respect to each mode of activity over time. \textbf{b. } Eigenvalue spectrum of optimized recurrent weights across $\lambda$. The contours indicate pseudospectra \citep{trefethenSpectraPseudospectraBehavior2005} at $\varepsilon = 0.03$. \textbf{c.} Effective rank of the recurrent weights across $\lambda$, as estimated by the number of singular values greater than 1e-2. Power law exponents are estimated over $\lambda \leq 1$. \textbf{d.} Participation ratio of activity versus $\lambda$. }
    \label{fig:2}
\end{figure}

Although this objective is nonconvex and difficult to optimize analytically in full generality, we can gain intuition for the classes of solutions it favors by considering its two limiting regimes. Where an exact characterization of the global optimum is unavailable, we construct explicit families of solutions that approach optimality in relevant limits and use them to illustrate the dynamical mechanisms associated with near-optimal solutions.
\subsection{Weight and activity efficiency favor distinct dynamical regimes}
In the limit $\lambda \rightarrow \infty$, we recover the minimum norm objective, given by $\mathcal{L}_{\lambda \rightarrow \infty}(\mW) \propto \|\mW\|^2_F  + \frac{1}{\|\vq(\tau)\|^2}$. For any delay period $\tau$, the optimal solution is then given by a readout-aligned, symmetric rank-1 spike of the form $W = \alpha \vc\vc^\top$, for some eigenvalue $\alpha \in [0,1)$. In particular, for $\tau \geq 1$, the optimal solution uses  marginally stable dynamics, with $\alpha \rightarrow 1^{-}$ (App. \ref{sec:weight_efficient}). Thus, a single persistent mode governed by line attractor-like dynamics \citep{seungHowBrainKeeps1996} is selected by the minimum norm objective. 

If we instead seek an activity-minimizing solution $\lambda \rightarrow 0$, the resulting objective is given by $\mathcal{L}_{\lambda \rightarrow 0}(\mW) = \left[\vq(\tau)^\top \mR^{-1} \vq(\tau) \right]^{-1}$. Intuitively, an activity-efficient solution should seek to transiently amplify activity around the readout time $\tau$, so as to not spend excess activity energy during the delay or after the readout. Moreover, the temporal resolution with which activity can be selectively amplified should be limited by the network size $N$. Indeed, we show that this activity-minimizing objective is lower bounded as $\mathcal{L}_{\lambda \rightarrow 0}(\mW) \geq \frac{\tau}{N}$ (App. \ref{sec:activity_efficient}). 

A natural candidate for achieving the highly temporally specific transient amplification necessary for saturating this bound is feedforward chain connectivity \citep{goldmanMemoryFeedbackNeural2009,ganguliMemoryTracesDynamical2008} of the form $\mA = -\mI + \mW = -a\mI + k\bm{\bm{\Gamma}}$, where $\bm{\bm{\Gamma}}_{ij} = \delta_{i+1,j}$, and $a$, $k >0$ represent the decay rate and transient gain, respectively (we have chosen a basis where the readout $\vc = \ve_1$, the first natural basis vector). With this choice of connectivity, the stimulus encoding is distributed and rotates through $N$ orthogonal directions before aligning with the readout direction. If we choose the decay rate to scale as $a = N/\tau$, and the amplification ratio $\rho = k/a$ to be large, we find that the activity-minimizing objective approaches $\mathcal{L}_{\lambda \rightarrow 0}(\mW) \approx \frac{\pi \tau}{N}$, thus achieving the bound up to a constant factor (App. \ref{sec:activity_efficient}). Thus, by increasing the length of the chain, this activity cost can be made arbitrarily small. We note that this construction is not unique; see \ref{sec:dirichlet} for a choice of $\mW$ based on high-dimensional, nonnormal oscillatory dynamics that can also saturate the bound, but up to a different constant factor.

We summarize in Fig. \ref{fig:1}b how various families of solutions differ in their respective weight and activity costs. Task-trained networks learn a recurrent mode of long timescale aligned with the input and readout directions,  allowing for high weight efficiency at the expense of activity efficiency (App. \ref{sec:weight_efficient}). On the other hand, networks with high-dimensional nonnormal feedforward structure can be constructed so as to be highly activity efficient but weight inefficient (App. \ref{sec:activity_efficient}). For solutions with unstructured recurrent weights $\mW_{ij} \sim \mathcal{N}(0, \sigma^2_W/N)$, both costs are comparably large, as such networks require large input vectors (and therefore large activity) so that the stimulus can still be read out at the proper scale by the readout time (App. \ref{sec:unstructured}). Finally, solving the constrained optimization in Eq. \ref{eqn:reduced} at different $\lambda$ yields the Pareto frontier of solutions for this task (Fig. \ref{fig:1}c,d).   

The optimal solutions selected by both limits of $\lambda$ are diametrically opposite extremes: the minimum norm solution uses a single persistent attracting mode, yet is highly activity inefficient, while the minimum activity solution is well-approximated by an arbitrarily high-rank chain with large amplification ratio, and thus is highly weight inefficient. These two extremes, while each pathological in their own way, are useful baselines for grounding the distinct inductive biases imposed by these normative objectives. At small but nonzero $\lambda$, it is natural to expect that the optimal solution would lie between these extremes, with a dimensionality controlled by $\lambda$. If we again consider a chain ansatz for $\mW$, but with an intermediate length $m \leq N$, the resulting activity term still approximately scales as $\left[\vq^\top (\mR+\lambda\mI)^{-1} \vq \right]^{-1} \sim \frac{\tau}{m}$, whereas the weight cost scales as $\|\mW\|_F^2 = m[(1-a)^2 + k^2] \sim \frac{m^3}{\tau^2}$. Balancing these costs yields an optimal chain length scaling as $m^\star \sim \lambda^{-1/4}\tau^{3/4}$.

\subsection{Activity-efficient solutions exhibit misaligned coding}
While these scaling exponents are not expected to be universal, and arise from a simplified ansatz for $\mW$, the conclusion that an extensive number of recurrent modes can be optimal to solve this simple memory task is robust to these modeling choices. Indeed, numerically optimizing the joint objective $\mathcal{L}(\mW)$ (App. \ref{sec:app_linear}) at different $\lambda$ yields a spectrum of increasingly high dimensional solutions as $\lambda$ approaches zero (Fig. \ref{fig:2}). For each numerically optimized solution $\mW$, we obtain an informative basis for understanding the properties of the solution via Arnoldi iteration, i.e., orthonormalizing the Krylov vectors $[\vc, \mW^\top \vc, \left(\mW^\top\right) ^2 \vc, \dots \left(\mW^\top\right) ^{m-1} \vc]$ generated by the readout direction, where $m$ is the effective rank of the recurrent weights. Viewed under this basis, we find that highly activity efficient networks can indeed be described as having nonnormal, amplifying chain-like connectivity, but with feedback connections rather than purely feedforward. Interestingly, despite the high dimensional nature of highly activity-constrained solutions, this structure is not immediately apparent from activity. In particular, the effective dimensionality of activity $\{\vx(t)\}$ remains $\mathcal{O}(1)$ across the whole range of $\lambda$ (Fig. \ref{fig:2}d). Continuing with the chain analogy, this could be explained by the fact that such solutions produce activity that is amplified exponentially as it propagates along the chain, so that the activity along the readout-aligned mode still remains the dominant contribution of explained variance. To quantify causal influence of each mode of activity $\vv _i$ on network output, we consider perturbations $\delta \vx(t) = \epsilon \vv _i$, and compute $\partial_\varepsilon y(\tau)|_{\epsilon = 0} = \vv_i^\top \vq(\tau-t)$ during the delay. 
Consistent with the feedforward chain analogy, we find that a dissociation between modes of activity that explain the most variance and those that maximally causally influence output emerges at low $\lambda$ (Fig. \ref{fig:2}a). Thus, such misaligned coding can arise simply from optimizing for activity efficiency. 

\subsection{Noise robustness competes with activity efficiency}\label{sec:noise}
\begin{figure}[t]
    \centering
    \includegraphics[width=\linewidth]{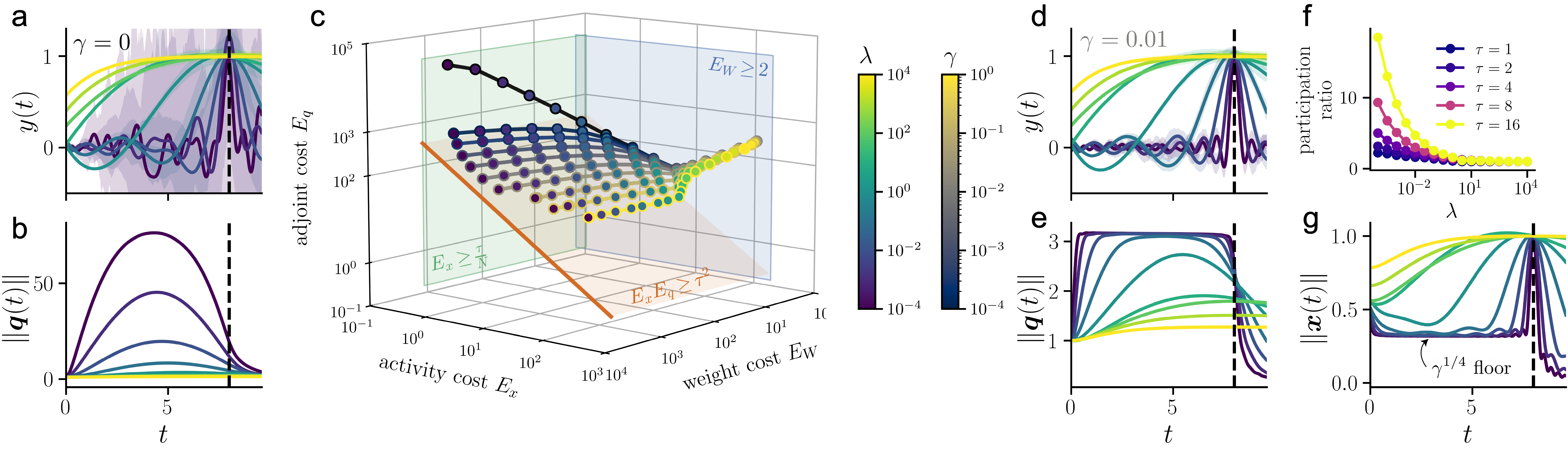}
    \caption{\textbf{Robustness to noise requires trading off against activity efficiency.} \textbf{a.} Output traces for optimized networks after dynamics are corrupted with process noise. Shaded regions indicate confidence intervals. \textbf{b.} Adjoint norm predicts noise sensitivity. \textbf{c.} Cost properties of networks optimized across both $\lambda$ and $\gamma$. The black trace corresponds to $\gamma=0$ (no adjoint penalty). Planes indicate theoretical bounds that constrain the solution space. \textbf{d-g.} Properties of networks optimized at nonzero adjoint cost $\gamma = 0.01$. \textbf{d.} Output traces under process noise no longer exhibit extreme noise sensitivity. Adjoint norms are constrained (\textbf{e}), and activity norms no longer approach zero during the delay (\textbf{g}). \textbf{f.} Effective dimensionality of activity now grows as $\lambda$ approaches zero. }
    \label{fig:3}
\end{figure}
Another common desideratum for biological circuits is robustness to noise. Unlike in feedforward networks, RNNs subject to process noise must contend with potential noise amplification arising from the successive application of the same recurrent weight matrix across timesteps. It is thus natural to ask whether optimizing for activity efficiency, which encourages temporally selective transient amplification of inputs, has the adverse consequence of increasing sensitivity to noise. Indeed, we find that networks optimized to solve the stimulus recall task in a highly activity efficient manner ($\lambda \ll 1$) were most sensitive to process noise (Fig. \ref{fig:3}a). 

If we augment the RNN's dynamics with white process noise as $\frac{d\vx}{dt} = -\vx + \mW\vx + \vb u(t) + \bm{\eta}(t)$, where  $\bm{\eta} \in \mathbb{R}^{N}$ satisfies $\mathbb{E}[\bm{\eta}(t)] = 0$ and $\mathbb{E}[\bm{\eta}(t)\bm{\eta}(t')^\top] = \sigma_\eta^2 \mI \delta(t-t')$, this induces noise in the output, with variance given by $\text{Var}_\eta[y(\tau)] = \sigma_\eta^2 \vc^\top \left(\int_0^{\tau} e^{\mA t}e^{\mA^\top t}dt\right) \vc = \sigma_\eta^2 \int_0^{\tau}\|\vq(t)\|^2 dt$. Thus, process noise induces noise in the readout with variance proportional to the integrated squared norm of the adjoint dynamics variable $\vq(t)$. To quantify sensitivity to noise in a task-agnostic (horizon-free) manner, it is therefore natural to consider an ``adjoint activity" given by the quantity 
\begin{equation}
    E_q = \int_0^{\infty} \|\vq(t)\|^2 dt,
\end{equation}
which measures the stationary variance along the readout due to isotropic process noise, and upper bounds the readout noise for any finite time horizon.

This apparent tradeoff between activity efficiency and noise robustness can be made precise by observing a standard conservation relation \citep{luchiniAdjointEquationsStability2014} between the time-reversed adjoint $\widetilde{\vq}(t) = \vq(\tau-t)$, defined for $t \in [0,\tau]$, and the activity variable $\widetilde{\vx}(t) = \vx(t) /z = e^{\mA t}\vb$. In particular, we have that $\frac{d}{dt} \widetilde{\vq}(t) ^\top \widetilde{\vx}(t) = \frac{d}{dt} (\vc^\top e^{\mA (\tau-t)}e^{\mA(t)}\vb) = 0$, and thus  $\widetilde{\vq}(t) ^\top \widetilde{\vx}(t) = \widetilde{\vq}(0) ^\top \widetilde{\vx}(0) = 1$ for all times $t \in [0,\tau]$, where the latter equality follows from the task constraint $g(\tau)=\vc ^\top e^{\mA \tau} \vb = 1$. Cauchy-Schwarz then yields the uncertainty relation-like bound 
\begin{equation}\label{eqn:uncertainty}
    E_x E_q \geq \tau^2,
\end{equation}
where $E_x = \int_0^{\infty}\|\widetilde{\vx}(t)\|^2 dt$ represents the average activity cost. In other words, minimizing activity necessarily demands that the noise sensitivity of the readout, as measured by the adjoint ``activity", grows to compensate, and vice versa. For the amplifying feedforward chain, this divergence of adjoint cost as the network becomes increasingly activity efficient can be shown explicitly (App. \ref{sec:app_noise}). 

We remark that this notion of noise sensitivity based on adjoint norm measures absolute noise scale, as opposed to a relative measure such as signal-to-noise ratio. This is appropriate for our setting, where the task constraint has fixed the scale of the ``signal" as $g(\tau) = 1$. Moreover, a mean-squared error style task objective of the form $\mathbb{E}_{z,\eta}[(y(\tau)-z)^2]$ decomposes additively as $\sigma_z^2\left(g(\tau)-1\right)^2 + \sigma_\eta^2 \int_0^{\tau}\|\vq(t)\|^2 dt$, thus making this a relevant notion of noise sensitivity for task-optimized RNNs as well. However, in settings where signal-to-noise ratio is optimized instead, amplifying nonnormal connectivity similar to that discussed here, paired with a judicious choice of input weights, can be optimal for extensive memory, as both signal and noise are amplified \citep{ganguliMemoryTracesDynamical2008}. 

If we now introduce the adjoint cost to the constrained optimization as 
\begin{equation}\label{eqn:problem2}
    \min_{\mW,\vb} E_x + \lambda E_W + \gamma E_q \qquad \text{s.t. } g(\tau) = 1,
\end{equation}
with weighting $\gamma$, we can damp the extreme sensitivity of solutions to process noise by tuning $\gamma$. In the limit of vanishing weight cost $\lambda \rightarrow 0$, optimal solutions hug the lower bound $E_x E_q \geq \tau^2$ (Fig. \ref{fig:3}e,g). Together with the approximately constant activity and adjoint norms during the delay observed numerically, this yields an activity scale $\|\vx(t)\| \approx \gamma^{1/4}$ throughout the delay (Fig. \ref{fig:3}g, App. \ref{sec:adjoint_floor}), in contrast to the vanishing delay activity observed before in Fig. \ref{fig:1}d. Moreover, modes of activity orthogonal to the readout direction become resolved rather than suppressed to infinitesimal magnitude (Fig. \ref{fig:supp_modes}), leading to an effective dimensionality of activity that grows with decreasing $\lambda$ and tracks the increase in recurrent rank, rather than staying $\mathcal{O}(1)$ (Fig. \ref{fig:3}f).

\section{Efficiency-robustness tradeoffs persist in nonlinear RNNs}
\begin{figure}[t]
    \centering
    \includegraphics[width=0.88\linewidth]{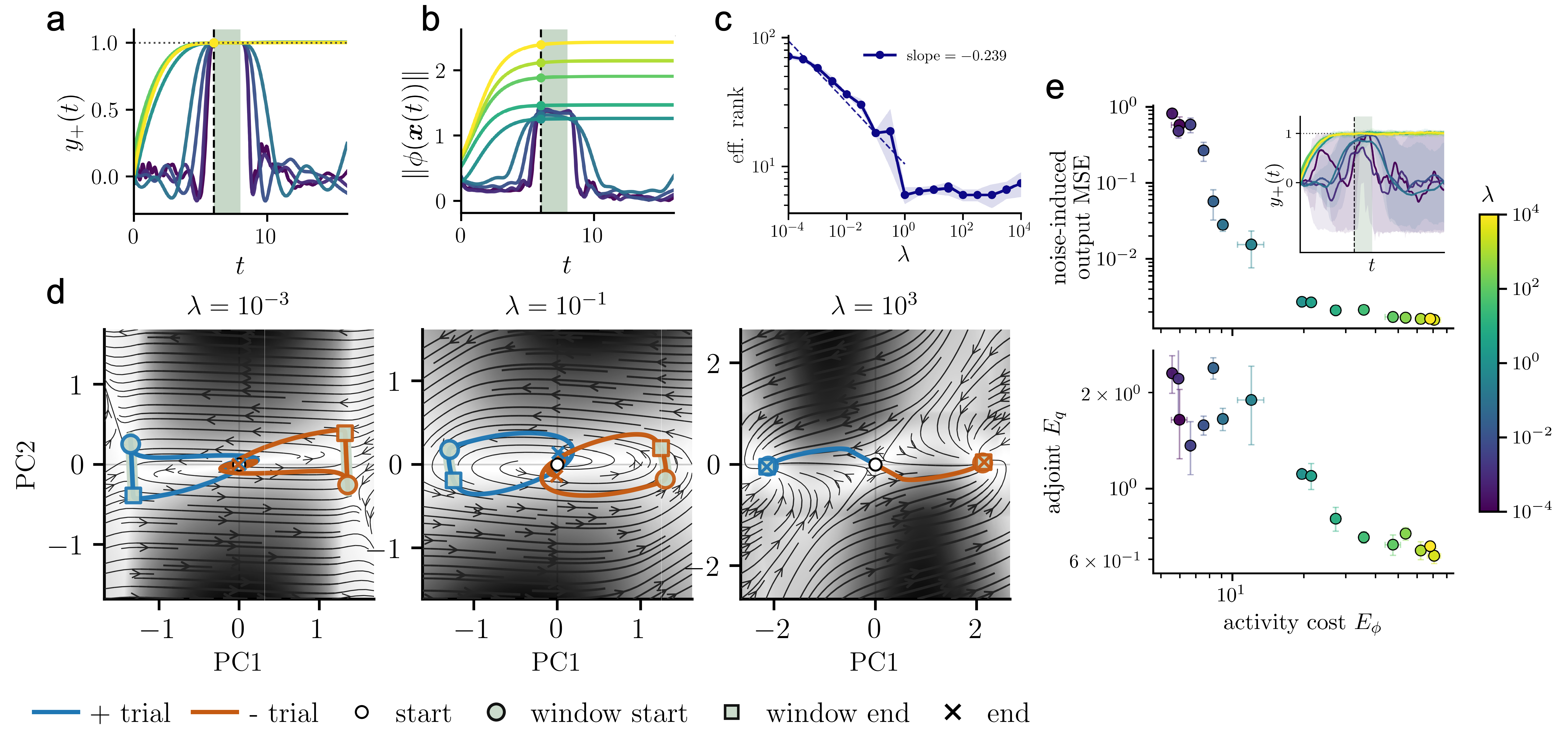}
    \caption{\textbf{Nonlinear RNNs exhibit similar changes in recurrent dynamics and noise sensitivity with increasing activity efficiency.} \textbf{a-b.} Output and activity over time across different $\lambda$, shown for a $+$ trial. The output is only constrained during the window highlighted in green. \textbf{c.} Effective rank of learned solutions across $\lambda$, averaged over five seeds. \textbf{d.} Flow fields projected onto the top two principal components of activity, with $+/-$ trial trajectories overlaid. Darkness of the background indicates flow velocity. \textbf{e.} Top: Output error due to injected process noise versus activity cost. Inset shows noised readouts for a $+$ trial. Each point is an average over seeds, with error bars indicating standard error of the mean. Bottom: Adjoint versus activity cost.}
    \label{fig:4}
\end{figure}
Thus far, we have considered linear RNNs subject to a temporally pointwise constraint on the network output, a tractable albeit restrictive setting. To test whether the qualitative tradeoffs identified above depend on these assumptions, we next consider nonlinear RNNs whose output is constrained over a continuous time window. Specifically, we train RNNs with $\phi =\tanh$ activation to solve a variant of the constrained optimization problem in Eq. \ref{eqn:problem}, but with the activity cost $E_\phi = \mathbb{E}_z\left[\int_0^{T} \|\phi(\vx(t))\|^2 dt\right]$ evaluated over a finite horizon $T$, the readout constraint $y(t) = \vc^\top \phi(\vx(t)) \approx z$ applied over $t \in [\tau, \tau + \xi]$, and the stimulus drawn discretely as $z \sim \text{Unif}(\{-1, +1\})$ (App. \ref{sec:app_nonlinear}). As it is difficult to solve this constrained optimization problem for nonlinear RNNs even numerically, we approximate doing so by adding the weight and activity costs as soft penalties, and assigning a large relative weighting to a soft task cost $\mathbb{E}_z\left[\int_\tau^{\tau+\xi} (y(t)-z)^2 dt\right]$ (App. \ref{sec:app_nonlinear}). Unlike linear RNNs, nonlinear RNNs can store memories of discrete stimuli across multiple stable fixed points. Indeed, we find that weight efficiency constrained networks find a low-rank solution that stores stimulus identity in distinct stable fixed points, as is typically observed in task-trained RNNs performing memory tasks \citep{sussilloOpeningBlackBox2013,maheswaranathanUniversalityIndividualityNeural2019,driscollFlexibleMultitaskComputation2024}. However, below a critical value of $\lambda$, solutions that rely on stable fixed points disappear (Fig. \ref{fig:4}). Instead, highly activity-constrained solutions learn high rank recurrent weights that produce choreographed transient trajectories, with a period of high activity aligned only to the readout window. Mirroring our findings in the simplified linear setting, we again find that the effective rank of trained solutions grows approximately as a power law with decreasing $\lambda$ (Fig. \ref{fig:4}). Thus, despite the richer dynamical repertoire available to nonlinear RNNs, activity efficient solutions still favor high rank recurrent weights. 

Next, we assess how network performance, as measured by mean-squared error of the output, degrades with injected neuronal noise. Consistent with our analysis of linear RNNs, we observe that greater activity efficiency is associated with greater sensitivity to noise (Fig. \ref{fig:4}e). While the conservation relation underlying Eq. \ref{eqn:uncertainty}, and hence an exact uncertainty bound, does not hold in general in the nonlinear setting, the cumulative squared norm of the adjoint variable $\tilde{\vq}(t) = \frac{\partial y(\tau)}{\partial \vx(t)}$ still scales roughly inversely to activity cost (Fig. \ref{fig:4}e). 

\begin{figure}[t]
    \centering
    \includegraphics[width=\linewidth]{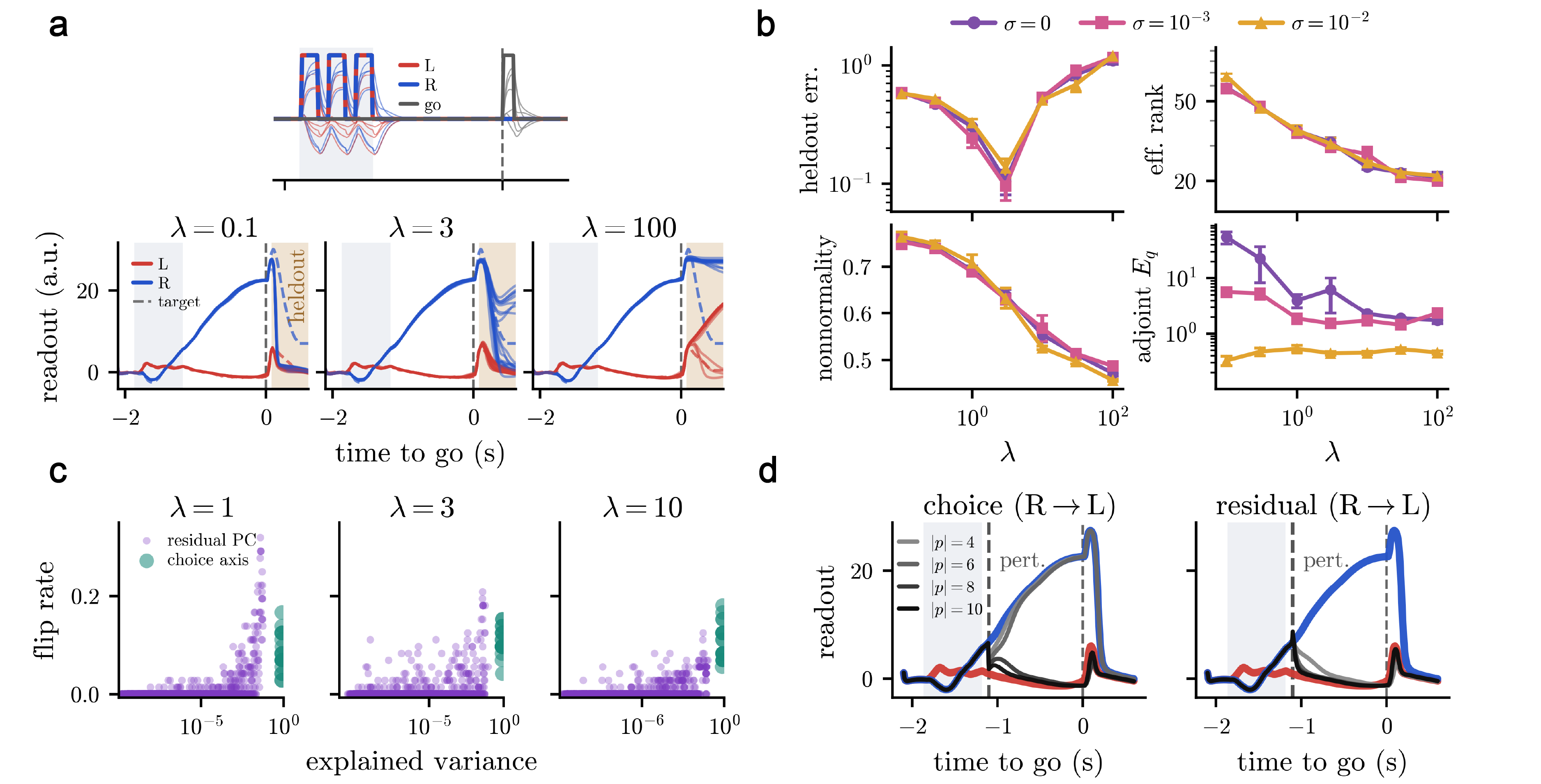}
    \caption{\textbf{An intermediate level of activity efficiency, rather than the minimum norm solution, recapitulates signatures of misaligned coding in anterior lateral motor cortex.} \textbf{a.} Top: Schematic of the task input. Low dimensional input pulses corresponding to left trial, right trial, and go cues are filtered by a fixed random reservoir before being passed as input to the network. Bottom: Readout traces across networks trained at different $\lambda$, shown for 20 random seeds. \textbf{b.} Properties of solutions trained at different process noise levels $\sigma$ and weight cost levels $\lambda$. \textbf{c.} Rate of flipping for perturbations aligned with residual principal components (purple) and choice axis (green), pooled over all 20 trained seeds. \textbf{d.} Example perturbation-induced responses for a representative network trained at $\lambda=1$ and $\sigma=1e-2$, shown for different perturbation magnitudes. }
    \label{fig:5}
\end{figure}
\subsection{Efficiency and robustness recapitulate misaligned coding in ALM}
Finally, we ask whether the same efficiency and robustness axes can recapitulate key dynamical properties of anterior lateral motor cortex during a memory-guided motor task. As in \citet{pereira-obilinovicNeuralDynamicsOutside2026}, we use networks with positive firing rates and learnable maximum firing rates per neuron:
\begin{align}
    \dot{\vx_i} &= -\vx_i + \sum_{j}\mW_{ij}\phi_j(\vx_j) +  \sum_{a}\mB_{ia}\bm{u}_a(t) + \vb_i^0 + \bm{\eta}_i\\[-6.5pt]
    y(t) &= \sum_{i} \vc_i\phi_i(\vx_i(t)),
\end{align}
with saturating nonlinearity $\phi_i(x) = \frac{r_i^{\text{max}}}{2}(\tanh(x)+1)$. Here, $\mB$ denotes the input weights, $\vb^0$ denotes trainable biases, and $\bm{\eta}$ is isotropic process noise. The task consists of ``left" or ``right" sample cues, a fixed delay, and then a subsequent ``go" cue, after which the network should produce a response corresponding to the sample cue that was observed (Fig. \ref{fig:5}a). These low-dimensional task inputs are first filtered through a fixed random reservoir to produce the high-dimensional input $\vu(t)$ that is passed into the RNN (App. \ref{sec:app_alm}). Here, rather than densely supervising the RNNs against recorded neural activity at the single-unit level, we ask if much weaker data supervision, along with some combination of the desiderata of weight efficiency, activity efficiency, and noise robustness could yield networks with qualitatively similar phenomenology as observed in \citet{pereira-obilinovicNeuralDynamicsOutside2026}. Specifically, we trained networks so that a readout of the firing rates reproduces the corresponding condition-averaged choice projection of neural activity recorded from anterior lateral motor cortex in mice performing the same task. In this manner, networks are constrained only through a one-dimensional, task-relevant projection of recorded activity. Importantly, this target displays a complex time-dependent profile with stimulus-dependent preparatory ramping, rather than simply tracking the stimulus. Among all such task-solving networks, we vary the weight-to-activity cost ratio $\lambda$ and the level of noise to determine their effect on the dynamical properties of resulting solutions (App. \ref{sec:app_alm}). We penalize recurrent and input weight norms as before, here $\|\mW\|_F^2 + \|\mB\|^2_F$, but do not penalize the biases. In addition, we fix the norm of the readout $\vc$ as $\|\vc\|=1$, but allow its direction to be trainable.  

To assess heldout predictivity, we supervised networks only up to a time just after the onset of the go cue, leaving the response period unconstrained. We find that both extremes of $\lambda$ led to networks that were poor predictors of the heldout targets (Fig. \ref{fig:5}b); highly activity-constrained networks produced readouts that decayed too quickly during the response period, whereas highly weight-constrained networks produced readouts that saturated and failed to decay at all. We also evaluated network performance on catch trials where the go cue was delayed or withheld, finding that training in the presence of process noise was necessary for networks to maintain stable choice-selective activity throughout the extended delay, consistent with experimental catch trials \citep{pereira-obilinovicNeuralDynamicsOutside2026}(Fig. \ref{fig:supp_catch}). We next assessed whether perturbations during the delay could flip the trained networks' readout trajectories from one outcome (i.e. left) to the other. For each network, we estimate sample, choice, and response directions by computing the average firing rate difference $\langle \vr^R(t) -\vr^L(t)\rangle$ between left and right trials over the sample, late-delay, and post-go cue periods, respectively (App. \ref{sec:app_pert}). We then defined residual directions as principal components of activity orthogonal to the subspace spanned by these task coding dimensions. Perturbing along both coding and residual dimensions, we find that decreasing $\lambda$, i.e., increasing activity efficiency, is associated with the emergence of residual directions with high causal influence, as measured by flip rate (Fig. \ref{fig:5}c,d, \ref{fig:supp_evflip}). Further, consistent with prior analyses, we find that such highly activity efficient solutions require high rank, nonnormal recurrent weights (Fig. \ref{fig:5}b), mirroring the nonnormal amplification implicated in the causal influence of residual directions found by \citet{pereira-obilinovicNeuralDynamicsOutside2026}. Overall, these results suggest that an intermediate level of activity efficiency, paired with robustness to process noise, is sufficient to recapitulate both the temporal structure of choice-related activity and the disproportionate causal influence of low-variance residual dimensions observed in ALM.
\section{Discussion}

Understanding how different forms of efficiency and robustness shape the space of solutions in ANNs is crucial for identifying and improving upon gaps in their analogy to biological neural circuits. In feedforward linear networks, constraints on the solution space imposed by minimizing total representation norm (activity) across layers are largely nested within those imposed by minimizing total weight norm \citep{braunNotAllSolutions2025}. Moreover, an objective that penalizes weight norm and/or representation norm arises naturally from optimizing for robustness (in the mean-squared error sense) of solutions to additive perturbations \citep{bishopTrainingNoiseEquivalent1995,braunNotAllSolutions2025, dorrellEfficientComputingTheory2026}. In contrast, we show that in recurrent neural networks, these desiderata of weight efficiency, activity efficiency, and noise robustness are in general in competition with each other, with each imposing distinct pressures on selected solutions. Although our analytical picture of these tradeoffs originates from a simplified setting of a linear RNN solving a simple memory task, we show that qualitatively similar tradeoffs arise even when dynamics are nonlinear, activations are bounded and positive, and targets have complex time-dependent profiles.

While solutions where task-relevant dynamics evolve in low-dimensional subspaces have been the focus of much of the computation-through-dynamics literature \citep{manteContextdependentComputationRecurrent2013,mastrogiuseppeLinkingConnectivityDynamics2018,valenteExtractingComputationalMechanisms2022}, we have identified activity efficiency as a biologically-relevant, interpretable scaling axis along which non-minimal, higher-dimensional solutions to even simple tasks may emerge. Further, such non-minimal solutions arising from optimizing for an intermediate level of activity efficiency appear to be more consistent with misaligned coding phenomena observed in anterior lateral motor cortex. 


Our work connects to prior studies linking recurrent dynamics supporting memory to underlying optimization principles \citep{ganguliMemoryTracesDynamical2008,limNoiseToleranceAttractor2011,stroudComputationalFoundationsDynamic2024,ritterEfficientWorkingMemory2025,stroudEffectsNoiseMetabolic2025, dorrellEfficientComputingTheory2026}. 
In the context of working memory, \citet{stroudOptimalInformationLoading2023} argued that inputs should be optimally loaded along directions that best take advantage of nonnormal structure in recurrent weights, and that such a scheme is consistent with neural recordings. \citet{ritterEfficientWorkingMemory2025} find that working memory networks optimized for signal-to-noise ratio and activity efficiency learn high-dimensional, nonnormal rotational dynamics, with qualitative similarities to the activity-efficient regime described here. Our results complement these studies by characterizing the interplay between weight, activity, and robustness costs in a setting where key aspects of their effects on recurrent weights and dynamics can be  described analytically. 

The present study has several limitations. First, we focus on simple memory tasks where only a single variable needs to be remembered to inform a later response. Correlations among stimuli and representational interference across time in sequence memory-style tasks \citep{sharmaTemporalSuperpositionFeature2025,dorrellEfficientComputingTheory2026} could substantially alter the nature of tradeoffs identified here. In addition, while we focus on particular forms of activity and weight efficiency, other desiderata may be more biologically relevant. For instance, due to homeostatic pressures, it may be more relevant to measure activity cost by the deviation from some baseline level of activity, rather than by overall scale \citep{chintaluriMetabolicallyRegulatedSpiking2023}. Finally, while we focused on how these different costs shape optimal solutions, we did not address the learning process that gives rise to such optimally efficient or robust solutions. Weight efficiency has a natural implementation as a local learning rule in the form of weight decay. It is less clear, however, how local learning rules would construct the high-rank, nonnormal connectivity favored by activity efficiency. Identifying plasticity rules that construct solutions favored by these less standard objectives is an important direction for future work.




\if\ificlrfinal

\subsubsection*{Acknowledgements}
We thank Jacob A. Zavatone-Veth and William Dorrell for insightful discussions and comments on a previous version of this manuscript, as well as members of the Pehlevan group for their feedback. W.Q. is supported by a Kempner Graduate Fellowship. C.P. is supported by an NSF CAREER Award (IIS-2239780), DARPA grants DIAL-FP-038 and AIQ-HR00112520041, the Simons Collaboration on the Physics of Learning and Neural Computation, and the William F. Milton Fund from Harvard University. This work has been made possible in part by a gift from the Chan Zuckerberg Initiative Foundation to establish the Kempner Institute for the Study of Natural and Artificial Intelligence.

\fi

\bibliography{iclr2027_conference}
\bibliographystyle{iclr2027_conference}

\clearpage 

\appendix
\counterwithin{figure}{section}
\section{Problem formulation} \label{sec:app_a}
\subsection{General formulation and reduction}
We first formulate the constrained optimization problem for a general distribution of readout times and a finite task-error tolerance. This makes clear which aspects of the reduction are generic, before specializing to the fixed-readout-time setting analyzed in the main text.

Consider a single-input single-output mapping governed by a linear RNN:
\begin{align}
    \dot{\vx} &= -\vx + \mW\vx + \vb u(t) \\
    y(t) &= \vc ^\top \vx(t)
\end{align}
where $\mW \in \mathbb{R}^{N\times N}$, $\vb \in \mathbb{R}^{N}$, and $\vc \in \mathbb{R}^N$.

On each trial, the network sees a pulse input $u(t) = z\delta(t)$, where $z \sim \mathcal{N}(0,1)$ represents an observed signal, and the network is simply required to reproduce the observed signal $z$ at a time $T \sim p(T)$. Crucially, on each trial, we do not explicitly constrain the readout $y(t)$ at any time other than the readout time $T$. We are interested in exploring the solution space of weights $\bm{\Theta} = [\mW,\vb, \vc]$ that allow the network to solve the task, up to an $\varepsilon$-slack in the expected squared-error task performance:
\begin{equation}
    \mathbb{E}_{z, T}\left[(y(T) - z)^2\right] \leq \varepsilon.
\end{equation}
As in the main text, we will fix a choice of readout vector $\vc$, with $\|\vc\| = 1$, and focus on choices of $\mW$ and $\vb$. We would like to characterize the constrained optimization problem  
\begin{align}
    \min_{\mW,\vb} \quad \mathbb{E}_z\left[\int_0^\infty \|\vx(t)\|^2 dt\right] + \lambda \left(\|\mW\|_F^2 + \|\vb\|^2\right) \\
    \text{s.t. } \mathbb{E}_{z, T}\left[(y(T) - z)^2\right] \leq \varepsilon.
\end{align}

where $\lambda$ is a parameter governing the relative importance of the activity and weight costs. 

Performing the expectation over $z$, the activity penalty becomes $\vb^\top \mR\vb$, where we define $\mR(\mW) = \int_0^\infty e^{\mA^\top t}e^{\mA t}dt$ and $\mA = -\mI+\mW$. Similarly, performing the expectation in the constraint yields
\begin{equation}
    \vb^\top \mQ\vb - 2\vq^\top\vb +1 \leq \varepsilon
\end{equation}
where we have defined $\mQ(\mW) = \mathbb{E}_T[e^{\mA^\top T}\vc \vc^\top e^{\mA T}]$ and $\vq(\mW) = \mathbb{E}_T[ e^{\mA^\top T}\vc]$.

Enforcing the constraint via Lagrange multiplier $\mu \geq 0$, we have:
\begin{equation}
    \mathcal{L}(\mW,\vb, \mu) = \vb^\top \mR \vb + \lambda \left(\|\mW\|^2_F + \|\vb\|^2\right) + \mu (\vb^\top \mQ \vb - 2 \vq^\top \vb +1 - \varepsilon).
\end{equation}

This objective is quadratic in $\vb$:
\begin{align}
    \mathcal{L}(\mW,\vb, \mu) = \vb^\top (\mR + \mu \mQ + \lambda \mI) \vb  -2 \mu \vq^\top \vb +\mu(1 - \varepsilon) + \lambda \|\mW\|^2_F
\end{align}
Minimizing in $\vb$, we find 
\begin{equation}
    \vb^* = \mu (\mR+\mu \mQ + \lambda \mI)^{-1}\vq,
\end{equation}
yielding the reduced objective 
\begin{equation}
    \mathcal{L}(\mW) = \lambda \|\mW\|^2_F + \sup_{\mu \geq 0} \left[\mu(1-\varepsilon) - \mu^2\vq^\top(\mR+\mu \mQ + \lambda \mI)^{-1}\vq\right].
\end{equation}

While simple to write down, it is difficult to make further analytical progress because the optimization over $\mu$ yields only an implicit fixed point equation for $\mu^*$. 

\subsection{Fixed readout time}\label{sec:reduced_problem}

We now specialize to the fixed readout-time setting of the main text, $p(T) = \delta(T - \tau)$, for which $\mQ = \vq\vq^\top$. By Sherman-Morrison, we have
\begin{align}
    \vq^\top(\mR+\mu \mQ + \lambda \mI)^{-1}\vq &= \vq^\top (\mR + \lambda \mI)^{-1}\vq - \mu\frac{(\vq^\top(\mR + \lambda \mI)^{-1}\vq)^2}{1 +\mu \vq^\top(\mR + \lambda \mI)^{-1}\vq} \\
    &= \frac{\vq^\top(\mR + \lambda \mI)^{-1}\vq}{1 + \mu \vq^\top(\mR + \lambda \mI)^{-1}\vq}.
\end{align}

Thus, defining $\alpha(\mW) = \vq^\top(\mR + \lambda \mI)^{-1}\vq$, the optimization over $\mu$ becomes straightforward:
\begin{align}
    \sup_{\mu \geq 0} \left[\mu(1-\varepsilon) - \frac{\mu^2 \alpha}{1 + \mu \alpha}\right] &= \sup_{\mu \geq 0} \left[\frac{\mu}{1+ \mu \alpha} - \varepsilon \mu \right]
\end{align}

Differentiating with respect to $\mu$ gives  $\mu^* = \frac{\varepsilon^{-1/2}-1}{\alpha}$, thus yielding
\begin{equation}
    \sup_{\mu \geq 0} \left[\frac{\mu}{1+ \mu \alpha} - \varepsilon \mu \right] = \frac{(1-\sqrt{\varepsilon})^2}{\alpha}.
\end{equation}
Thus, the reduced objective is 
\begin{equation}
    \mathcal{L}(\mW) = \lambda \|\mW\|_F^2 + \frac{(1-\sqrt{\varepsilon})^2}{\vq^\top(\mR + \lambda \mI)^{-1}\vq}.
\end{equation}
Note that this is similar to the objective in the main text, but with the latter term rescaled by a factor $(1-\sqrt{\varepsilon})^2$. Therefore, at least for the fixed readout-time constraint, permitting some small slack $\varepsilon > 0$ does not appreciably change the space of solutions.  

Similarly, we can simplify the optimal input vector as
\begin{equation}
    \vb^* = \frac{(1-\sqrt{\varepsilon})(\mR+\lambda \mI)^{-1}\vq}{\vq^\top(\mR + \lambda \mI)^{-1}\vq}.
\end{equation}
To recover the setup of the main text, we set $\varepsilon = 0$:
\begin{equation}
    \mathcal{L}(\mW) = \lambda \|\mW\|_F^2 + \frac{1}{\vq^\top(\mR + \lambda \mI)^{-1}\vq}.
\end{equation}
\section{Weight and activity efficiency}\label{sec:app_B}
\subsection{Unstructured recurrent weights}\label{sec:unstructured}
Suppose we select and fix recurrent weights randomly as $\mW_{ij} \sim \mathcal{N}\left(0,\frac{\sigma^2_W}{N}\right)$, fix readout weights $\vc$ such that $\|\vc\| = 1$, and choose input weights $\vb$ as 
\begin{align}
    \vb^* &= \text{argmin}_{\vb} \|\vb\| \\
    \text{s.t.} \quad g(\tau) &= 1.
\end{align}

This corresponds to a reservoir-like regime. The input weights are then given as $\vb^* = \frac{\vq(\tau)}{\|\vq(\tau)\|^2}$, where $\vq(\tau) = e^{\mA ^\top \tau}\vc$. We would then like to estimate the weight and activity costs $E_W$ and $E_x$ under this choice of parameters. Conveniently, all quantities of interest can be estimated from order parameters of the dynamical mean field theory description of random RNNs, which yields closed-form expressions here as we are considering linear RNNs. We give a brief self-contained version of the calculation here, which follows that of \citet{bordelonDynamicsLearningIntegrate2026}. 

Observe that $\vq(t)$ obeys the dynamics
\begin{equation}
    \frac{d\vq}{dt} = -\vq + \mW^\top \vq,
\end{equation}
subject to initial condition $\vq(0)= \vc$. Squaring these dynamics yields 
\begin{equation}
    (1+\partial _t )(1+\partial_{t'}) \vq(t) \vq(t')^\top =  \mW ^\top \vq(t) \vq(t')^\top \mW.
\end{equation}

Defining the autocorrelation function $C(t,t') = \frac{1}{N}\langle \vq(t)^\top \vq(t') \rangle $, we have that 
\begin{align}
    (1+\partial_t)(1+\partial_{t'}) C(t,t') &= \frac{1}{N}\Tr(\langle \mW ^\top \vq(t) \vq(t')^\top \mW\rangle) \\
    &= \frac{1}{N} \langle \vq(t')^\top  \mW \mW^\top \vq(t)\rangle \\
    &\approx \frac{1}{N} \langle \vq(t')^\top \langle \mW \mW^\top \rangle\vq(t)\rangle \\
    &= \sigma_W^2 C(t,t').
\end{align}

This PDE has closed-form solution 
\begin{equation}
    C(t,t') = C(0,0)e^{-t-t'}I_0\left(2\sqrt{\sigma_W^2 tt'}\right),
\end{equation}

where $I_0$ is a modified Bessel function. Thus, 
\begin{align}
    \|\vb^*\|^2 &= \frac{1}{NC(\tau, \tau)} \approx 2\sqrt{\pi\sigma_W \tau} e^{2(1-\sigma_W)\tau},
\end{align}
where we have used the leading asymptotics $I_0(x) \sim \frac{e^x}{\sqrt{2\pi x}} $, and the initial condition $C(0,0) = \frac{1}{N}\vc^\top \vc = \frac{1}{N}$. The same mean-field description of the statistics of the corresponding activity variable $\widetilde{\vx}(t) = e^{\mA t}\vb^*$ is difficult to apply directly because the initial condition $\vb^*$ is strongly dependent on the particular choice of $\mW$, but we nonetheless can obtain a bound corresponding to the activity resulting from a generic initial condition of the same norm (valid under mean-field assumptions): 
\begin{align}
    E_x =\int_0^{\infty} \|\widetilde{\vx}(t)\|^2 dt &\ \gtrsim N \int_0^{\infty} C(t,t)|_{C(0,0) = \frac{1}{N}\|\vb^*\|^2} dt \\
    &=  \frac{\|\vb^*\|^2}{2\sqrt{1-\sigma_W^2}}. 
\end{align}

Thus, both the total weight cost $E_W = N\sigma^2_W + \|\vb^*\|^2$ and the activity cost $E_x$ scale at least as large as $\sim e^{2(1-\sigma_W)\tau}$.

One can also choose input weights $\vb$ to be minimum norm under the activity metric, i.e.
\begin{align}
    \vb^* &= \text{argmin}_{\vb} \|\vb\|_{\text{act}} \\
    \text{s.t.} \quad g(\tau) &= 1,
\end{align}
where $\|\vb\|^2_{\text{act}} = \vb^\top \mR \vb$, corresponding to a minimizer $\vb^* = \frac{\mR^{-1}\vq(\tau)}{\vq(\tau)^{\top}\mR^{-1}\vq(\tau)}$. Numerically, this yields solutions with moderately smaller activity cost and moderately larger weight cost (Fig. \ref{fig:1}b).  
\subsection{Gradient descent from small initialization}\label{sec:gd}
Here, we show that gradient descent from small initialization recovers the solution of minimum norm, provided that we clamp the recurrent weights $\mW$ to be stable. Initialize parameters as $\mW_{ij}^0, \vb^0_i, \vc^0_i \sim \mathcal{N}\left(0, \frac{\sigma^2_W}{N}\right) $. The loss can be expressed as 
\begin{equation}
    L = \mathbb{E}_{z}[(y(\tau)-z)^2] = \sigma^2_z(g(\tau)-1)^2.
\end{equation}
For convenience, we set $\sigma^2_z = 1$. 

We then have the gradient flow dynamics
\begin{align}
    \frac{d}{ds}\mW^s &= - \eta (g(\tau)-1)\int_0^\tau \vq(t) \widetilde{\vx}(\tau-t)^\top dt\\
    \frac{d}{ds}\vb^s &= -\eta (g(\tau)-1)\vq(\tau) \\
    \frac{d}{ds}\vc^s &= -\eta (g(\tau)-1)\widetilde{\vx}(\tau) ,
\end{align}

where we have absorbed the factor of $2$ into the learning rate $\eta$. At early training times $s$, assuming $\sigma_W \ll 1$, we have, up to first-order in $\sigma_W$, that 
\begin{align}
    \frac{d}{ds}\mW^s &\approx \eta \tau e^{-\tau} \vc^s (\vb^s)^\top \\
    \frac{d}{ds}\vb^s &= \eta e^{-\tau}\vc^s \\
    \frac{d}{ds}\vc^s &= \eta e^{-\tau}\vb^s
\end{align}

where we used the early-training-time approximations $\vq(t) \approx e^{-t}\vc^s + \mathcal{O}(\sigma_W^2)$ and $\widetilde{\vx}(t) \approx e^{-t}\vb^s + \mathcal{O}(\sigma_W^2)$. As in \citep{bordelonDynamicsLearningIntegrate2026}, $\vb^s$, $\vc^s$, and $\mW^s$ quickly align to the common direction $\vr \propto \frac{1}{2}(\vb^0 + \vc^0)$, yielding the reduced-order description $\mW ^s \approx w(s) \vr \vr^\top$, $\vb^s \approx b(s) \vr$, $\vc^s \approx c(s) \vr$ for $\|\vr\|=1$. Plugging this ansatz into the full gradient flow then yields the reduced flow
\begin{align}
    \frac{d}{ds}w(s) &=- \eta \left(b(s)c(s)e^{(w(s)-1)\tau}-1\right)b(s)c(s) e^{(w(s)-1)\tau}\tau \\
    \frac{d}{ds} b(s) &= - \eta \left(b(s)c(s)e^{(w(s)-1)\tau}-1\right) e^{(w(s)-1)\tau}c(s) \\
    \frac{d}{ds}c(s) &=  - \eta \left(b(s)c(s)e^{(w(s)-1)\tau}-1\right) e^{(w(s)-1)\tau}b(s)
\end{align}

Observe that this flow obeys the conservation law  $\frac{d}{ds} (b(s)^2 - c(s)^2) = 0$. Thus, assuming a balanced initialization $b(0)=c(0)$, we can take $c(s) = b(s)$. This yields
\begin{align}
        \frac{d}{ds}w(s) &=- \eta \left(b(s)^2e^{(w(s)-1)\tau}-1\right)b(s)^2 e^{(w(s)-1)\tau}\tau \\
    \frac{d}{ds} b(s) &= - \eta \left(b(s)^2e^{(w(s)-1)\tau}-1\right) e^{(w(s)-1)\tau}b(s).
\end{align}
We then have that 
\begin{equation}
\frac{d}{ds} b(s)^2 = 2b(s) b'(s) = -2 \eta \left(b(s)^2e^{(w(s)-1)\tau}-1\right) e^{(w(s)-1)\tau}b(s)^2.
\end{equation}
Thus, 
\begin{equation}
    \frac{d w(s)}{d (b(s)^2)} = \frac{\tau}{2},
\end{equation}
implying that $w(s) - \frac{\tau}{2}b(s)^2$ is conserved. Since we assume small initialization, this yields the relation
\begin{equation}\label{eqn:conserved}
    w(s) = \frac{\tau}{2}b(s)^2,
\end{equation}
which holds over all training times $s$. The endpoint of training is given by the condition for solving the task $g(\tau)=1$, giving 
\begin{align}\label{eqn:endpoint}
    b^2 e^{(w-1)\tau} = \frac{2}{\tau}we^{(w-1)\tau} = 1.
\end{align}
However, for $\tau > 2$, the corresponding $w$ satisfying Eq. \ref{eqn:endpoint} requires $w>1$, yielding unstable recurrent weights. Thus, if we replace $w$ with a clamped variable $w_{c}(s) = \min(w(s),1)$, then $w_c(s)$ obeys the relation $w_c(s) = \frac{\tau}{2}b(s)^2$ only up to the first time $s_c$ such that $w_c(s_c) = 1$. Once $w_c$ is clamped, the input/output norm obeys the dynamics 
\begin{equation}
    \frac{d}{ds}b(s)^2 = -2 \eta (b(s)^2 - 1)b(s)^2
\end{equation}
from initial condition $b(s_c)^2 = \frac{2}{\tau}$. Since this is now a logistic ODE, we have that $\lim_{s \rightarrow \infty} b(s)^2 = 1$. 

\begin{figure}
    \centering
    \includegraphics[width=0.8\linewidth]{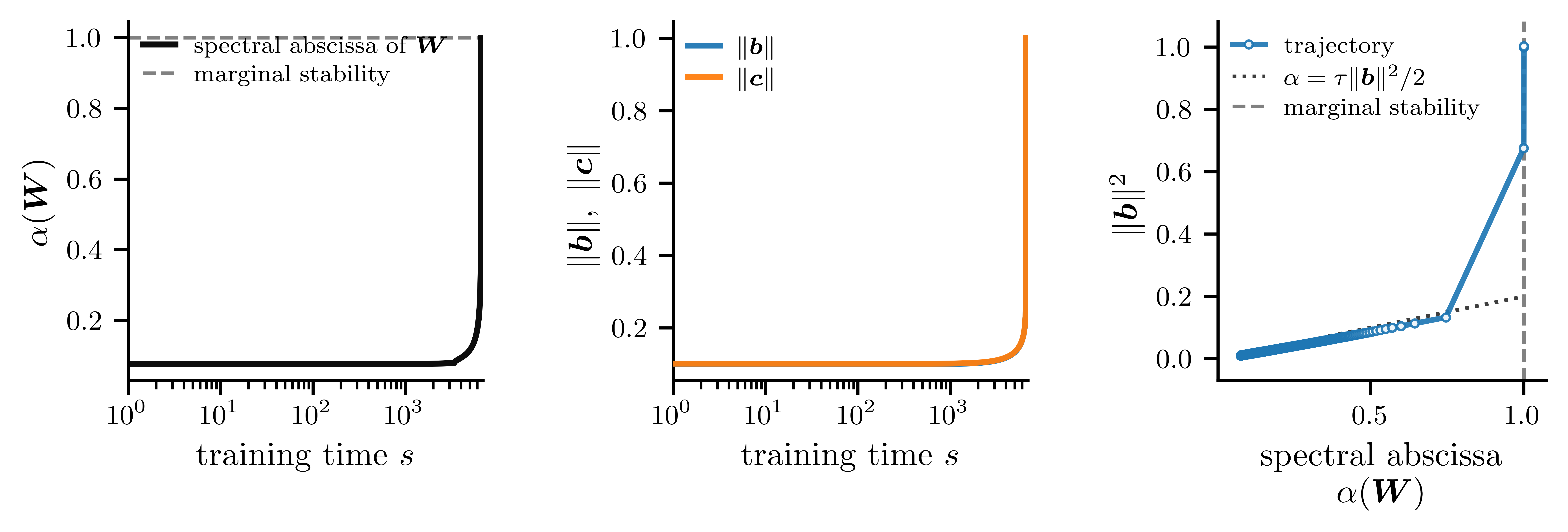}
    \caption{\textbf{Learning the stimulus recall task via gradient descent on the MSE loss from small initialization.} Left: Spectral abscissa of the recurrent weights over training time. Middle: Norms of the input and output weights over training time. Right: Learning trajectory in the spectral abscissa versus input weight norm plane closely tracks the conservation relation in Eq. \ref{eqn:conserved}.} 
    \label{fig:placeholder}
\end{figure}
Thus, with stability clamping of the recurrent weights, GD from small initialization yields the minimum norm solution: the recurrent weights, input, and output weights align along the same single direction, with marginally stable recurrent dynamics. In the absence of stability clamping, the weights still align in this manner, but the recurrent weights become unstable.  
\subsection{Weight-constrained regime}\label{sec:weight_efficient}

Taking $\lambda$ to be large, we find:

\begin{equation}\label{eqn:obj_large_lam}
    \mathcal{L}(\mW) = \lambda \left(\|\mW\|_F^2 + \frac{1}{\|\vq\|^2} + \frac{1}{\lambda}\frac{\vq^\top \mR \vq}{\|\vq\|^4} + O\left(\frac{1}{\lambda^2}\right)\right).
\end{equation}
We can bound $\|\vq\|$ as:
\begin{align}
\|\vq\| = e^{-\tau}\|e^{\mW^\top \tau}\vc\| &\leq e^{-\tau}\|e^{\mW^\top \tau}\|_2\|\vc\|  \\
&\leq e^{\tau (\|\mW\|_2-1)} \\
&\leq e^{\tau (\|\mW\|_F-1)}.
\end{align}
Thus, we have a lower bound 
\begin{equation}
    \mathcal{L}/\lambda \geq \alpha^2 + e^{2\tau(1 - \alpha)} + O(1/\lambda),
\end{equation}
where $\alpha = \|\mW\|_F$. Taking $W = \alpha \vc\vc^\top$ saturates this lower bound. The optimal $\alpha$ is then given by the minimizer of $f(\alpha) = \alpha^2 + e^{2\tau (1-\alpha)}$. Observing that $f'(0) <0$ and $f'(1) = 2(1-\tau)$, for $0 < \tau < 1$, the minimizer is the solution to $f'(\alpha^*) = 2(\alpha^* -\tau e^{-2(\alpha^*-1)\tau}) = 0$. Thus, $W \propto \vc\vc^\top$ is globally optimal for $\tau \leq 1$. However, for $\tau > 1$, the minimizer requires $\alpha^* > 1$, which leads to divergent activity, and thus is disallowed even at infinitesimal effective activity cost strength $\lambda ^{-1}$. 

We will now prove that even for $\tau > 1$, provided we restrict $\mW$ to be stable, a global optimum (in the infimum sense) is given by $W \propto \vc\vc^\top$, and in particular by $\mW_0 = \alpha\vc\vc^\top$ for $\alpha \rightarrow 1^{-}$. 

Defining 
\begin{equation}
    \tilde{\mathcal{L}} = \|\mW\|^2_F + \frac{1}{\|\vq\|^2},
\end{equation}
we have that $\tilde{\mathcal{L}}(\mW_0) = 2$. 

It thus suffices to show that for all stable $\mW$ (i.e., $\max(\Re(\lambda(\mW)) < 1)$, $\tilde{\mathcal{L}}(\mW) \geq 2$. 

For $\|\mW\|_F \geq \sqrt{2}$, the claim immediately follows from the nonnegativity of $1/\|\vq\|^2$. 

For $\|\mW\|_F < 1$, since the lower bound of $\tilde{\mathcal{L}}$, $\alpha^2 + e^{2\tau(1-\alpha)}$, is monotonically decreasing in $\alpha$ over $0 < \alpha < 1$ for $\tau > 1$, the optimal loss achievable for $0 < \alpha <1$ is given by $\alpha \rightarrow {1}^-$, yielding the desired result.

Thus it remains to show that the claim still holds for $1 < \|\mW\|_F < \sqrt{2}$ and $\tau > 1$. The claim in this case is nontrivial because nonnormal amplification can in general make $\|\vq\|$ grow beyond $1$, and standard exponential growth bounds on $\|\vq(t)\|$ are too loose. The proof of this subcase is not terribly illuminating, but we give it here for completeness. 

From $\|\vq\| = \|e^{\mA^\top \tau}\vc\| \leq \|e^{\mA \tau}\|_2$, it suffices to prove the stronger statement that 
\begin{equation}
    \|e^{\mA\tau}\|_2^2 \leq \frac{1}{2-\|\mW\|_F^2}.
\end{equation}
We first prove this statement in the special case of real upper-triangular 2x2 matrices of the form $\mW = \begin{pmatrix}
    a & b \\
    0 & d
\end{pmatrix}$, for $a, d < 1$ (stability criterion), and $1 \leq \|\mW \|^2_F =a^2+b^2+d^2< 2$. We will later show that all other cases can be reduced to this case by induction.

We have that 
\begin{equation}
    e^{(-\mI + \mW)\tau} = \begin{pmatrix}
        e^{(a-1)\tau} & b C \\
        0 & e^{(d-1)\tau}
    \end{pmatrix},
\end{equation}
where $C = \int_0^{\tau} e^{(a-1)(\tau - t)}e^{(d-1)t} dt$. Then, to show that $\|e^{(-\mI+\mW)\tau}\|_2^2 \leq \frac{1}{2-\|\mW\|^2_F}$, we would like to prove 
\begin{equation}
    \frac{1}{2-\|\mW\|^2_F}\mI - \left(e^{(-\mI + \mW)\tau}\right)^\top e^{(-\mI + \mW)\tau} \succeq \bm{0}.
\end{equation}
Since this is 2x2, this reduces to the conditions $e^{2(a-1)\tau} \leq \frac{1}{\Delta}$ (top left) and the determinant condition
\begin{equation} \label{eqn:det}
    \left(1-\Delta e^{2(a-1)\tau}\right)\left(1-\Delta e^{2(d-1)\tau}\right) \geq \Delta b^2C^2,
\end{equation}
where we introduced the shorthand $\Delta = 2-\|\mW\|^2_F$. The top left condition follows trivially from observing that $0 <\Delta \leq 1$, and thus $\frac{1}{\Delta} \geq 1$ and $e^{2(a-1)\tau} \leq 1$. For the determinant condition, we can rewrite
\begin{align}
    (1-\Delta e^{2(a-1)\tau})(1-\Delta e^{2(d-1)\tau}) &= \left(1- \Delta e^{(a+d-2)\tau}\right)^2 - \Delta \left(e^{(a-1)\tau}-e^{(d-1)\tau}\right)^2.
\end{align}
We also have the explicit integral $C  = \frac{1}{a-d}\left( e^{(a-1)\tau} - e^{(d-1)\tau}\right)$. Thus, to show \ref{eqn:det}, we would like to show that
\begin{equation} \label{eqn:det2}
   \left( 1 - \Delta e^{(a+d-2)\tau}\right)^2 \geq \Delta ((a-d)^2 + b^2) C^2.
\end{equation}
To prove this, it will be convenient to first show that \begin{equation}
    \Delta ((a-d)^2 + b^2) \leq (a+d-2)^2.
\end{equation}
From $\Delta = 2 - a^2 - b^2 - d^2$, we have that 
\begin{align} 
    \Delta ((a-d)^2 + b^2)  &= \Delta (2- 2ad - \Delta),
\end{align}
thus giving
\begin{align}
    (a+d-2)^2 - \Delta ((a-d)^2 + b^2) &=  ((a-1)+(d-1))^2 - \Delta (2-2ad-\Delta) \\
    &= ((a-1)+(d-1)+\Delta)^2 + 2\Delta (1-a)(1-d)  \\ &\geq 0.
\end{align}
From \ref{eqn:det2}, it then suffices to prove 
\begin{equation} \label{eqn:det3}
    1 - \Delta e^{(a+d-2)\tau} \geq (2-a-d)C.
\end{equation}
We have the pointwise bound 
\begin{equation}
    ((1-a) + (1-d))e^{(a-1)(t-s)}e^{(d-1)s} \leq (1-a)e^{(a+d-2)(t-s)}+(1-d)e^{(a+d-2)s},
\end{equation}
which integrating from $0$ to $\tau$, yields
\begin{equation}
    (2-a-d)C \leq 1 - e^{(a+d-2)\tau}.
\end{equation}
Since $0 \leq \Delta <1 $, we have that $1 - e^{(a+d-2)\tau} \leq 1 - \Delta e^{(a+d-2)\tau}$, thus showing \ref{eqn:det3}, completing the proof for the upper triangular 2x2 case. 

For generic $\mW$, we have the complex Schur decomposition $\mW = \mU \mT \mU^*$, for $\mU, \mT \in \mathbb{C}^{N \times N}$, $\mU^* \mU = \mU\mU^* = \mI$, and $\mT$ upper triangular. Define $\mT^R$ such that $T^R_{ii} = \Re{T_{ii}}$, $T^R_{ij} = |T_{ij}|$ for $i < j$. Observe that $\|e^{\mA \tau}\|_2 = \|e^{(-\mI + \mT) \tau}\|_2$. We then would like to show that it is sufficient to consider only the corresponding real upper triangular matrix $\mT^R$ by bounding 
\begin{equation} \label{eqn:bound_real}
    \|e^{(-\mI+\mT)\tau}\|_2 \leq \|e^{(-\mI+\mT^{R})\tau}\|_2.
\end{equation}
To show this, we will show a stronger pointwise statement that $|e^{(-\mI+\mT)\tau}\vv|  \leq e^{(-\mI+\mT^R)\tau}|\vv|$, for any $\vv \in \mathbb{C}$, where $|\cdot|$ operates entrywise. Defining $\vx(t) = e^{(-\mI+\mT)t}\vv$ and $\vy(t) = e^{(-\mI+\mT^R)t}|\vv|$, we have 
\begin{align}
    x_i(t) &= e^{(-1 + T_{ii})t}v_i + \int_0^{t}e^{(-1 + T_{ii})(t-s)}\sum_{i < j}T_{ij}x_j(s) ds.
\end{align}
We will now use backwards induction starting from $i = N$ down to $i=1$ to show this pointwise bound. For the base case $i=N$, $x_N(t) = e^{(-1+T_{NN})t}v_N$, so 
\begin{equation}
    |x_N(t)| = e^{(-1 + \Re T_{NN})t}|v_N| = y_N(t).
\end{equation}
For general $i$, we have
\begin{align}
    |x_i(t)| &\leq e^{(-1 + \Re T_{ii})t}|v_i| + \int_0^{t}e^{(-1 + \Re T_{ii})(t-s)}\sum_{i < j}|T_{ij}||x_j(s)| ds \\
    &\leq ^{(-1 + \Re T_{ii})t}|v_i| + \int_0^{t}e^{(-1 + \Re T_{ii})(t-s)}\sum_{i < j}|T_{ij}|y_j(s) ds \\
    &= y_i(t),
\end{align}
where the latter inequality follows from the induction hypothesis. Thus, we have shown that $|e^{(-\mI+\mT)\tau}\vv|  \leq e^{(-\mI+\mT^R)\tau}|\vv|$. 
Therefore,
\begin{align}
    \|e^{(-\mI + \mT) \tau}\vv \|_2  &\leq \|e^{(-\mI + \mT^R) \tau}|\vv| \|_2 \\
    &\leq  \|e^{(-\mI + \mT^R) \tau}\|_2\|\vv \|_2.
\end{align}
Taking the supremum over $\vv \neq 0$ yields the desired bound \ref{eqn:bound_real}. Combining this with the observation $\|\mT^{R}\|_F \leq \|\mT\|_F = \|\mW\|_F$, we have that for any $\mW$ obeying $1 < \|\mW\|_F < \sqrt{2}$ and $\tau > 1$, it suffices to show the bound for the corresponding real upper triangular matrix $\mT^R$,
\begin{equation} \label{eqn:real_tri}
    \left\|e^{(-I+\mT^{R})\tau}\right\|_2^2 \leq \frac{1}{2-\left\|\mT^R\right\|^2_F}.
\end{equation}

We will now prove \ref{eqn:real_tri} by induction on dimension $N$. For the base case $N=1$, let $r = T_{11}^R$. Recall that $r < 1$ (stability) and $r^2 < 2$ (Frobenius norm bound). Since $\tau > 1$, 
\begin{equation}
    e^{2(-1+r)\tau} \leq e^{2(-1+r)} \leq \frac{1}{2 - r^2},
\end{equation}
where the second inequality follows from $2-r^2 \leq 1 + 2 (1-r) \leq e^{2(1-r)}$.

For $N = k$ for some $k >1$, we can write 
\begin{equation}
    \mT^R = \begin{pmatrix}
        \mB & \vv \\
        \bm{0}^\top & d
    \end{pmatrix},
\end{equation}
for $\mB \in \mathbb{R}^{(N-1) \times (N-1)}$ an upper triangular matrix, $\vv \in \mathbb{R}^{N-1}$, and $d \in \mathbb{R}$. Let $B = \|\mB\|_F$ and $v = \|\vv\|_F$. Defining $\vh(\tau) = \int_0^{\tau} e^{(-\mI_{N-1}+\mB)(\tau-t)}\vv e^{(d-1)t} dt$, we have that 
\begin{align}
    \left\|e^{(-I+\mT^{R})\tau}\right\|_2 &= \left\|\begin{pmatrix}
        e^{(-\mI_{N-1}+\mB)\tau} & \vh(\tau) \\
        \bm{0}^\top & e^{(d-1)\tau}
    \end{pmatrix}\right\| \\
    &\leq \left\|\begin{pmatrix} \label{eqn:compression}
        \left\|e^{(-\mI_{N-1}+\mB)\tau}\right\|_2 & \|\vh(\tau)\|_2 \\
        0 & e^{(d-1)\tau}
    \end{pmatrix}\right\|.
\end{align}
We now consider two cases. If $B < 1$, $\|e^{(-\mI_{N-1}+\mB)\tau}\|_2 \leq e^{(B-1)\tau}$. Thus, $\|\vh(\tau)\| \leq v\int_0^{\tau}e^{(B-1)(\tau-t)}e^{(d-1)t}dt$, thus yielding the bound
\begin{align}
    \left\|e^{(-I+\mT^{R})\tau}\right\|_2 &\leq \left\|\begin{pmatrix}
        e^{(B-1)\tau} & v \int_0^{\tau} e^{(B-1)(\tau - t)}e^{(d-1)t} dt \\
        0 & e^{(d-1)\tau}
    \end{pmatrix}\right\| \\
    &= \left\|\exp\left(\left(-\mI_2 + \begin{pmatrix}
        B & v \\
        0 & d
    \end{pmatrix}\right)\tau\right)\right\| \\
    &\leq \frac{1}{\sqrt{2-\|\mT^R\|^2_F}},
\end{align}
where the last inequality follows from the 2x2 upper triangular bound. 
In the case $B > 1$, we have (from \ref{eqn:compression})
\begin{align}\label{eqn:induction_step}
    \left\|e^{(-I+\mT^{R})\tau}\right\|_2 &\leq \left\| \begin{pmatrix}
    \frac{1}{\sqrt{2-B^2}} & \frac{v }{\sqrt{2-B^2}}\int_0^{\tau}e^{(d-1)t}dt \\
    0 & e^{(d-1)\tau}
    \end{pmatrix}\right\|,
\end{align}
where we have applied the induction hypothesis on $\mB$, as well as the bound 
\begin{align}
    \|\vh(\tau)\| &\leq \int_0^\tau \|e^{(-\mI_{N-1} + \mB)(\tau-t)}\|_2 \|\vv\|e^{(d-1)t}dt \\
    &\leq \frac{v}{\sqrt{2-B^2}}\int_0^\tau e^{(d-1)t}dt,
\end{align}
where for $\tau - t \geq 1$, we use the induction hypothesis directly, whereas for $0 \leq \tau - t < 1$, we use $\|e^{(-\mI + \mB)(\tau-t)}\| \leq e^{(B-1)(\tau -t)} \leq e^{B-1} \leq \frac{1}{\sqrt{2-B^2}}$.
Observe that we can factor the matrix on the RHS of \ref{eqn:induction_step} as 
\begin{equation}
    \begin{pmatrix}
        \frac{1}{\sqrt{2-B^2}} & 0 \\
        0 & 1
    \end{pmatrix}
    \begin{pmatrix}
        1 & v\int_0^{\tau}e^{(d-1)t}dt \\
        0 & e^{(d-1)\tau}
    \end{pmatrix} = \begin{pmatrix}
        \frac{1}{\sqrt{2-B^2}} & 0 \\
        0 & 1
    \end{pmatrix}\exp\left(\left(-\mI_2 + \begin{pmatrix}
        1 & v \\
        0 & d
    \end{pmatrix}\right)\tau\right).
\end{equation}
Thus, for $B >1$, we have the bound
\begin{align}
    \left\|e^{(-I+\mT^{R})\tau}\right\|_2^2 \leq \frac{1}{2-B^2} \cdot \frac{1}{2-(1+v^2+d^2)}, 
\end{align}
which we would like to upper bound by $\frac{1}{2 - \|\mT^R\|^2_F} = \frac{1}{2 - B^2 - v^2 - d^2}$. 

Observe that 
\begin{align}
    (2-B^2)(1-v^2-d^2) - (2 - B^2 - v^2 -d^2) 
    &= (v^2+d^2)(B^2-1) \\
    &\geq 0,
\end{align}
and thus $(2-B^2)(1-v^2-d^2) \geq (2 - B^2 - v^2 -d^2) $, proving the desired claim. 
\subsection{Activity-constrained regime}\label{sec:activity_efficient}

The limit $\lambda \rightarrow 0$ is a bit more difficult to analyze in full generality. Our approach will be to show that a particular structured choice of $W$ (but not necessarily the only choice) can achieve arbitrarily small cost $\mathcal{L}$ when $\lambda = 0$, and then derive how the optimal $W$ within this  family of solutions adjusts to small but nonzero $\lambda$. For $\lambda = 0$, the objective reduces to 
\begin{equation}
    \mathcal{L}(\mW) = \frac{1}{\vq^\top \mR^{-1}\vq}.
\end{equation}

We first establish a lower bound on this term, and then show by construction a network that can achieve the lower bound, up to a constant factor. 

Observe that $\vq^\top \mR^{-1}\vq = \vc^\top e^{\mA\tau}R^{-1}e^{\mA^\top \tau}\vc = \vc^\top \mM(\tau) \vc$, where $\mM(\tau) = e^{\mA\tau}\mR^{-1}e^{\mA^\top \tau}$. Note that $\mM(\tau) \succeq 0$. We can also show that $\mM(\tau)$ is monotonically decreasing in $\tau$ in the Loewner order:
\begin{equation}
    \frac{d}{d\tau} \mM = e^{\mA \tau}( \mA\mR^{-1} + \mR^{-1}\mA^\top)e^{\mA^\top \tau}.
\end{equation}

Since $\mR$ satisfies the Lyapunov equation $\mA^\top \mR + \mR\mA = -\mI$, left and right multiplying by $\mR^{-1}$ yields the identity $\mA\mR^{-1} + \mR^{-1}\mA^\top = -\mR^{-2}$. Thus, we have that 
\begin{equation}
    \frac{d}{d\tau} \mM = -e^{\mA\tau}\mR^{-2}e^{\mA^\top \tau} \preceq 0.
\end{equation}

Moreover, we have the exact identity $\int_0^{\infty} \Tr(\mM(t))dt = \int_0^{\infty}\Tr (\mR^{-1}e^{\mA^\top t}e^{\mA t}) dt = \Tr(\mI) = N$. 

Thus, we have the bounds
\begin{align}
 \tau \Tr(\mM(\tau)) \leq \int_0^{\tau} \Tr(\mM(\tau)) dt \leq N,
\end{align}
yielding $\Tr(\mM(\tau)) \leq \frac{N}{\tau}$. Finally, this yields
\begin{equation}
    \vc^\top \mM(\tau) \vc \leq \lambda_{max}(\mM(\tau)) \leq \frac{N}{\tau}.
\end{equation}
Thus, 
\begin{equation}
    \frac{1}{\vq^\top \mR^{-1}\vq} \geq \frac{\tau}{N}. 
\end{equation}

The idealized solution in this limit is to produce a spike of activity along $\vc$ precisely at the readout time $\tau$. Since this kind of sudden transient amplification requires extremely nonnormal dynamics to execute, we consider a choice of recurrent weights of the homogeneous feedforward chain
\begin{equation}
    \mA = \mW-\mI = -a\mI + k\bm{\Gamma},
\end{equation}
where $\bm{\Gamma}_{ij} = \delta_{i+1,j}$, and $a, k > 0$ represent the decay rate and nonnormal amplification strength, respectively. 

It is convenient to write $\mA=\mD\mJ\mD^{-1}$, where
\begin{equation}
        \mD=\operatorname{Diag}(1,\rho^{-1},\rho^{-2},\ldots,\rho^{-(N-1)}),
    \quad
    \mJ=a(-\mI+\bm{\Gamma}) .
\end{equation}
Indeed, since \(\mD\bm{\Gamma} \mD^{-1}=\rho\bm{\Gamma}\), we have
\[
    \mD\mJ\mD^{-1}
    =
    -a\mI+a\rho\bm{\Gamma}
    =
    -a\mI+k\bm{\Gamma}
    =
    \mA.
\]
Then,
\[
    \vq^\top (\mR+\lambda \mI)^{-1}\vq
    =
     \tilde{\vq}^\top( \tilde{\mR}+\lambda \mD^2)^{-1}\tilde{\vq},
\]
where
\[
    \tilde \vq=e^{\mJ^\top\tau}\ve_1,
    \quad
    \tilde \mR=\int_0^\infty e^{\mJ^\top t}D^2e^{\mJ t}\,dt.
\]
Taking \(\rho\) to be large and \(\lambda\) to \(0\), we have
\[
    \tilde \mR
    =
    \int_0^\infty
    e^{\mJ^\top t}
    \left(
        \ve_1\ve_1^\top
        +
        \rho^{-2}\ve_2\ve_2^\top
        +
        \rho^{-4}\ve_3\ve_3^\top
        +\cdots
    \right)
    e^{\mJ t}\,dt
\]
\[
    =
    \int_0^\infty e^{-2at}\vf(t)\vf(t)^\top\,dt
    +O(\rho^{-2}),
\]
where
\[
    \vf(t)=
    \left(
        1,\,
        at,\,
        \frac{(at)^2}{2!},\,
        \ldots,\,
        \frac{(at)^{N-1}}{(N-1)!}
    \right)^\top.
\]
Similarly,
\[
    \tilde \vq=e^{-a\tau}\vf(\tau).
\]
Observe that, to leading order in $\rho^{-2}$, \(\tilde \mR_{ij}=\langle f_i(t),f_j(t)\rangle_a\)
is a Gram matrix of polynomials \((at)^i/i!\) up to degree \(N-1\)
under the weighted inner product
\[
    \langle f,g\rangle_a
    =
    \int_0^\infty e^{-2at}f(t)g(t)\,dt .
\]
Thus, letting
\[
    \bm{\Phi}(t)=(\phi_0,\phi_1,\ldots,\phi_{N-1})^\top
\]
be the orthogonal basis for this inner product space, we have that $\tilde{\mR}^{-1}$
\[
    \tilde \vq^\top \tilde \mR^{-1}\tilde \vq
    =
    e^{-2a\tau}
    \sum_{i=0}^{N-1}\phi_i(\tau)^2
    +O(\rho^{-2}).
\]
For the weight \(e^{-2at}\), this basis is given by the Laguerre polynomials
\[
    \phi_n(t)=\sqrt{2a}\,L_n(2at).
\]
Taking \(N\) to be proportional to \(a\tau\) as
\[
    x=2a\tau=Nz,
\]
we have from asymptotics of Laguerre kernels that \citep{forresterAsymptoticFormDensity2005}
\[
    e^{-x}\sum_{n=0}^{N-1}L_n(x)^2
    =
    \frac{1}{2\pi}\sqrt{\frac{4-z}{z}}
    +O(1/N).
\]
Thus,
\[
    \tilde \vq^\top \tilde \mR^{-1}\tilde \vq
    \approx
    2a\cdot
    \frac{1}{2\pi}\sqrt{\frac{4-z}{z}}
    =
    \frac{N}{2\pi\tau}\sqrt{z(4-z)}.
\]
This is maximized at \(z=2\), yielding
\[
    L(a^*)\big|_{\lambda=0,\rho\to\infty}
    =
    \frac{\pi\tau}{N},
\]
with
\[
    a^*=\frac{N}{\tau}.
\]
We now justify that taking $\rho \rightarrow \infty$ is indeed optimal for minimizing this activity cost term. We can rewrite the denominator as 
\begin{align}
    \tilde{\vq}^\top (\tilde{\mR}+\lambda \mD^2)^{-1} \tilde{\vq} &= \sup_{\vu \in \mathbb{R}^N} \frac{(\vu^\top \tilde{\vq})^2}{\vu^\top (\tilde{\mR}+\lambda \mD^2)\vu}.
\end{align}
Observing that $\vu^\top \tilde{\vq} = e^{-a\tau}\sum_{i=0}^{N-1}u_i \frac{(a\tau)^i}{i!}$, we can thus define a polynomial $p_u(s)$ of degree $N-1$ such that $\vu^\top \tilde{\vq} = e^{-a\tau}p_u(a\tau)$. Using the expansion $\tilde{\mR} = \sum_{i=0}^{N-1} \int_0^{\infty}e^{-2at}\rho^{-2i}\vf^{(i)}(t){\vf^{(i)}(t)}^\top dt$, where $\cdot^{(i)}$ denotes the $i$th derivative, we can then pass to an optimization over polynomials $p$ of degree at most $N-1$:
\begin{equation}\label{eqn:poly}
    \tilde{\vq}^\top (\tilde{\mR}+\lambda \mD^2)^{-1} \tilde{\vq} = \sup_{p: deg(p) \leq N-1} \frac{e^{-2a\tau}p(a\tau)^2}{\sum_{i=0}^{N-1}\int_0^{\infty}\rho^{-2i}p^{(i)}(at)^2 dt + \lambda \sum_{i=0}^{N-1}\rho^{-2i}p^{(i)}(0)^2}.
\end{equation}
Since the denominator is nonincreasing in $\rho$, it follows that the activity cost term $ (\tilde{\vq}^\top (\tilde{\mR}+\lambda \mD^2)^{-1}\tilde{\vq})^{-1}$ is nonincreasing in $\rho$. Thus, taking $\rho \rightarrow \infty$ is optimal when $\lambda \rightarrow 0$. 

The resulting objective cost improves towards zero as $1/N$; an infinitely long chain with infinitely large decay rate and amplification factor therefore achieves zero activity cost. This is pathological, so we are interested in understanding what the optimal amplifying chain is when $\lambda$ is small but nonzero. In particular, there should be an optimal chain length $N^* < \infty$ which should decay with $\lambda$. 

Heuristically, for small but nonzero $\lambda$, $\rho \rightarrow \infty$ is disallowed, but $k >a$ must still hold for transient growth in state norm to occur. Assuming the optimal decay rate retains the scaling $a \sim N/\tau$ found above, $k^*$ must therefore scale at least as large as $N/\tau$, yielding that the Frobenius penalty scales as $\|\mW\|^2_F = N ((1-a)^2 + k^2) \sim \frac{N^3}{\tau^2}$. This yields 
\begin{equation}
    \mathcal{L} \sim C_1\frac{\tau}{N} + \lambda C_2\frac{N^3}{\tau^2},
\end{equation}
for some $\mathcal{O}(1)$ constants $C_1$ and $C_2$. Thus, for small but nonzero $\lambda$, there is a finite optimal number of modes $N$ that minimizes the loss, which scales like $N^* \sim \lambda^{-1/4}\tau^{3/4}$. We emphasize that the exact exponents derived here are not universal, and are likely specific to the particular ansatz for $W$ analyzed here.

\subsection{Alternative activity-efficient construction based on oscillatory dynamics}\label{sec:dirichlet}
Another natural way of constructing a highly activity-efficient solution is to construct a linear dynamical system that implements a Dirichlet kernel approximation of a temporally narrow spike around the readout time. In particular, we seek to use
\begin{equation}
    D_n(x) = \sum_{k=-n}^n e^{ikx} = 1 + 2 \sum_{k=1}^n \cos(kx),
\end{equation}
which approaches a periodic delta function as $D_\infty (x) = 2\pi\sum_{k=-\infty}^\infty \delta(x +2\pi k)$. Thus, choosing weights such that the filter satisfies $g(t) = \frac{1}{N}e^{-a(t-\tau)}D_n(\omega (t-\tau))$ with frequency $\omega = \frac{\pi}{\tau}$ yields a spike of magnitude $1$ at $t=\tau$, with periodic (but exponentially damped) spikes reoccurring at $t=3\tau,5\tau$, etc. 

Let $N = 2n+1$. If we choose recurrent weights $\mW_0$ such that $\mA_0 = -\mI + \mW_0$ takes the form
\begin{equation}
    \mA_0 = \begin{pmatrix}
        -a &  \\
         & -a & -\omega &  &    &  \\
         & \omega & -a &  &  &   \\
         &  &   & -a & -2\omega &  \\
         &  &   & 2\omega & -a &  \\
          &  &  &  &  & \ddots \\
          & & & & & & -a & -n\omega \\
          & & & & & & n\omega & -a
    \end{pmatrix},
\end{equation}
where $a$ is the decay rate, and $\omega$ is the base frequency. 

Let $\mA_0^k$ denote the $k^{th}$ 2x2 diagonal block. 

With readout weights $\vc = \begin{bmatrix}
    c^{(0)} & (\vc^{(1)})^\top & \dots & (\vc^{(n)})^\top 
\end{bmatrix}^\top$
where $\vc^{(k)} \in \mathbb{R}^2$ for $1 \leq k \leq n$, and $c^{(0)} \in \mathbb{R}$

and input weights  $\vb_0 = \begin{bmatrix}
    b^{(0)} & (\vb^{(1)})^\top & \dots & (\vb^{(n)})^\top 
\end{bmatrix}^\top$ with components defined likewise. 

Then, the filter implemented by the system is given by 
\begin{align}
    g(t) &= \vc^\top e^{\mA_0 t}\vb_0 = e^{-at}c^{(0)}b^{(0)} + \sum_{k=1}^{n} (\vc^{(k)})^{\top} e^{\mA^k_0 t}\vb^{(k)} \\
    &= e^{-at} \left (c^{(0)}b^{(0)} + \sum_{k=1}^n (\vc^{(k)})^{\top} \begin{pmatrix}
        \cos(k\omega t) & \sin(k\omega t) \\
        -\sin(k\omega t) & \cos(k\omega t) \\
    \end{pmatrix}\vb^{(k)} \right)
\end{align}

Then, it is easy to see that choosing $b^{(0)} = \frac{e^{a\tau}}{\sqrt{N}}$, $\vb^{(k)} = \frac{e^{a\tau} (-1)^k}{\sqrt{N}}\begin{pmatrix}
    \sqrt{2} & 0
\end{pmatrix}^\top $ and $c^{(0)} = \frac{1}{\sqrt{N}}$, $\vc^{(k)} = \frac{1}{\sqrt{N}}\begin{pmatrix}
    \sqrt{2} & 0
\end{pmatrix}^\top$ yields 
\begin{align}
    g(t) &= \frac{e^{-a(t-\tau)}}{N}\left(1 + 2\sum_{k=1}^n \cos(k\omega t) (-1)^k\right).
\end{align}
Choosing $\omega = \frac{\pi}{\tau}$ then yields
\begin{align}
    g(t) &= \frac{e^{-a(t-\tau)}}{N}\left(1 + 2 \sum_{k=1}^n \cos(k\omega (t-\tau)) \right) \\
    &= \frac{e^{-a(t-\tau)}}{N} D_n(\omega (t-\tau)),
\end{align}
which has the desired property that $g(\tau) =1$. 

However, as stated, this construction uses a normal dynamics matrix $W$, and thus cannot achieve transient amplification necessary to stay activity efficient. In particular, the total activity is given by 
\begin{align}
    E_x = \int_0^{\infty} \|\widetilde{\vx}_0(t)\|^2 dt &= \int_0^{\infty}\|e^{\mA_0t}\vb_0\|^2 dt  \\
    &= \frac{1}{N}\int_0^{\infty}e^{-2a(t-\tau)}\left(1 + 2\sum_{k=1}^n (\cos ^2 (k\omega t) + \sin^{2}(k\omega t))\right) \\
    &= \frac{e^{2a\tau}}{2a}.
\end{align}

This quantity is minimized at $a^* = \frac{1}{2\tau}$, thus yielding a total activity of $E_x^* = e\tau$, which does not improve with $N$.

To improve the activity efficiency of this solution, we apply a similarity transform $\mS = I + (\kappa-1) \vc\vc^\top$, such that $\mA = \mS\mA_0\mS^{-1}$, together with $\vb = \frac{1}{\kappa}\mS\vb_0$. This stretches activity along the readout direction, relative to other orthogonal directions, by a factor $\kappa$, yielding a filter $g(t)$ identical to that implemented by $(\vc, \mW_0, \vb_0)$. The total activity under these transformations then becomes
\begin{align}
    E_x &= \frac{1}{\kappa^2}\int _0^{\infty} \vb_0^\top e^{\mA_0^\top t} \mS^\top \mS e^{\mA_0t}\vb_0 dt \\
    &= \frac{1}{\kappa^2}\int _0^{\infty} \vb_0^\top e^{\mA_0^\top t} (\mI + (\kappa^2-1)\vc\vc^\top) e^{\mA_0t}\vb_0 dt \\
    &= \frac{1}{\kappa^2}\int_0^{\infty}\|\widetilde{\vx}_0(t)\|^2 dt + \left(1 - \frac{1}{\kappa^2}\right)\int_0^{\infty}g(t)^2 dt \\
    &= \frac{1}{\kappa^2} \frac{e^{2a\tau}}{2a} + \left(1 - \frac{1}{\kappa^2}\right)\int_0^{\infty}g(t)^2 dt.
\end{align}

Thus, for large $\kappa$, the total activity under this transformation approaches $\int_0^{\infty} g(t)^2 dt$, and is therefore concentrated along the readout direction.

The readout-aligned activity integral can then be simplified as (Mathematica)
\begin{align}
    \int_0^{\infty} g(t)^2 dt &= \frac{1}{N^2}\int_0^{\infty}e^{-2a(t-\tau)}D_n^2(\omega(t-\tau)) dt \\
    &= \frac{\tau e^{2a\tau}}{N}\left[\frac{1}{2a\tau} + 4a\tau\sum_{m=1}^{N-1} \frac{\left(1 - \frac{m}{N}\right)(-1)^m}{4(a\tau)^2+m^2 \pi^2} \right] \\
    &= \frac{\tau e^{2a\tau}}{N}\left[ A_N - \frac{1}{N}B_N \right],
\end{align}
where $A_N = \frac{1}{2a\tau} + 4a\tau \sum_{m=1}^{N-1} \frac{(-1)^m}{4(a\tau)^2+m^2 \pi^2}$ and $B_N = 4a\tau\sum_{m=1}^{N-1} \frac{m(-1)^m}{4(a\tau)^2+m^2 \pi^2}$.

Observe that $\lim_{N \rightarrow \infty} A_N = \frac{1}{\sinh(2a\tau)}$ exactly, and thus $A_N = \frac{1}{\sinh(2a\tau)} + \mathcal{O}(N^{-2})$. In addition, we have that $B_N = \mathcal{O}(1)$ as $B_N$ is convergent. Thus, we have the leading asymptotics 
\begin{align}
    \int_0^{\infty} g(t)^2 dt &= \frac{\tau e^{2a\tau}}{N \sinh(2a \tau)} + \mathcal{O}\left( \frac{1}{N^2}\right) \\
    &= \frac{2\tau}{N (1-e^{-4a\tau})} +  \mathcal{O}\left( \frac{1}{N^2}\right).
\end{align}

Estimating the optimal $a^\star$ would require more careful treatment of the subleading terms. For illustrative purposes however, setting $a = \frac{1}{2\tau}$ yields $\int_0^{\infty} g(t)^2 dt \approx \frac{2\tau}{N(1-e^{-2})}$. 

Thus, in the limit $\kappa \rightarrow \infty$, this nonnormal oscillatory construction also achieves an activity cost $E_x = \mathcal{O}\left(\frac{\tau}{N}\right)$, demonstrating that the activity-efficient scaling derived above does not rely specifically on feedforward chain connectivity. Nonetheless, like the feedforward chain, it relies on transient amplification arising from high-rank, nonnormal recurrent weights. 



\section{Noise robustness}\label{sec:app_noise}
As in Sec. \ref{sec:noise}, we consider isotropic process noise added to the dynamics of variance $\sigma^2$ injected into the dynamics:
\begin{equation}
    \frac{d\vx}{dt} = -\vx + \mW\vx + \vb u(t) + \bm{\eta}(t),
\end{equation}
where $\mathbb{E}[\bm{\eta}(t)] = 0$ and $\mathbb{E}[\bm{\eta}(t)\bm{\eta}(t')^\top] = \sigma^2 I \delta(t-t')$. Then, the readout $y(\tau)$ satisfies 
\begin{align}
    \mathbb{E}_\eta[y(\tau)] &= (\vc^\top e^{\mA\tau}\vb) z \\
    \text{Var}_\eta[y(\tau)] &= \sigma^2 \vc^\top \left(\int_0^{\tau} e^{\mA t}e^{\mA^\top t}dt\right) \vc = \sigma^2 \int_0^{\tau}\|\vq(t)\|^2 dt
\end{align}

If we define $\widetilde{\vq}(t) = \vq(\tau - t)$ for $t \in [0,\tau]$, the more standard convention for the adjoint dynamics variable that evolves backwards in time, it is easy to see that 
\begin{equation}
    \frac{d}{dt} \widetilde{\vq}(t)^\top \vx(t)  = \frac{d}{dt}\vc^\top e^{\mA(\tau-t)}e^{\mA t}\vb = \frac{d}{dt} \vc^\top e^{\mA\tau}\vb = 0.
\end{equation}

Thus, combining this with the readout gain constraint $\vc^\top e^{\mA\tau}\vb = \widetilde{\vq}(0)^\top \widetilde{\vx}(0) = 1$, we have the conservation law that 
\begin{equation}
    \widetilde{\vq}(t)^\top \widetilde{\vx}(t) = 1
\end{equation}
for  all $t \in [0, \tau]$. 
Thus, minimizing activity norm necessarily requires $\widetilde{\vq}$ to grow in norm to compensate. More concretely, from Cauchy-Schwarz, we have:
\begin{align}
    \left(\int_0^{\tau} \widetilde{\vq}(t)^\top \widetilde{\vx}(t) dt\right)^2 \leq \left(\int_0^{\tau} \|\widetilde{\vx}(t)\|^2 dt \right) \left(\int_0^{\tau} \|\widetilde{\vq}(t)\|^2 dt\right) \leq E_x E_q. 
\end{align}
Thus, $E_x E_q \geq \tau^2$. 
\subsection{Noise robustness of the amplifying chain}
For the concrete ansatz of the homogeneous feedforward chain $\mA = -a\mI + k\bm{\Gamma}$, the asymptotics of $E_q$ can be explicitly computed:
\begin{equation}
    E_q = \int_0^{\infty}\|\vq(t)\|^2 dt
\end{equation}
Using that $q_i(t) = e^{-at}\frac{(kt)^i}{i!}$, we have that 
\begin{align}
    \|\vq(t)\|^2 &= e^{-2at}\sum_{j=0}^{N-1}\frac{(kt)^{2j}}{(j!)^2} 
\end{align}

Performing the integral then yields
\begin{equation}
    E_q = \frac{1}{2a}\sum_{j=0}^{N-1}\binom{2j}{j}\left(\frac{\rho^2}{4}\right)^j.
\end{equation}

Using the approximation that $\binom{2j}{j} \sim \frac{4^j}{\sqrt{\pi j}}$, we have that 
\begin{equation}
    E_q \sim \frac{1}{2a}\sum_{j=0}^{N-1} \frac{\rho^{2j}}{\sqrt{\pi j}}.
\end{equation}
For $\rho > 1$, this expression diverges exponentially fast in $N$. 

\subsection{Adjoint cost sets an activity scale}\label{sec:adjoint_floor}
In the limit of vanishing weight cost $\lambda \rightarrow 0$, the activity and adjoint costs approach the lower bound $E_xE_q \approx \tau^2$. Under this assumption, we have that $E_x + \gamma E_q = E_x + \gamma \frac{\tau^2}{E_x}$, which then yields $E_x^* = \gamma ^{1/2} \tau$ and $E_q^* = \gamma^{-1/2}\tau$. If we then take $\|\widetilde{\vx}(t) \|$ and $\|\vq(t)\|$ to be constant throughout $[0,\tau]$, as is observed numerically, we then obtain $\|\widetilde{\vx}(t)\| \approx \gamma^{1/4}$ and $\|\vq(t)\| \approx \gamma^{-1/4}$ for $t \in [0,\tau]$. 
\section{Numerical experiments and training details}\label{sec:numerics}
Details for running numerical experiments in this paper are stated below. All code will be made publicly available upon acceptance. 
\subsection{Linear RNN experiments}\label{sec:app_linear}
To numerically optimize the objective in Eq. \ref{eqn:reduced}, we used the dissipative Hamiltonian parameterization \citep{mehlStabilityRadiiLinear2016} of the effective recurrent weights $\mA = -\mI + \mW$, given by $$\mA = (\mJ-\mR_0)\mQ,$$ where $\mJ$ is skew-symmetric, $\mR_0 \succ \bm{0}$, and $\mQ \succ \bm{0}$, thus enforcing Hurwitz stability by construction. This was necessary to avoid numerical issues when solving the Lyapunov equation $\mA ^\top \mR + \mR \mA = -\mI$ for the Gramian $\mR$ appearing in the activity cost using a Lyapunov solver. 

Optimization was performed in JAX with the L-BFGS-B optimizer. We use this optimizer because standard first-order methods converged much more slowly. We used networks of size $N=30$ (Fig. \ref{fig:1}) or $N=50$ (Figs. \ref{fig:2}, \ref{fig:3}). These network sizes were chosen for efficiency of the optimization process, and were selected to be large enough to accommodate the largest effective rank observed across the $\lambda$ and $\tau$ values considered. 

\subsection{Nonlinear RNN experiment (continuous readout window task)}\label{sec:app_nonlinear}
As described in the main text, we use $\phi=\tanh$ RNNs of the form 
\begin{equation}
    \dot{\vx} = -\vx + \mW \phi(\vx(t)) + \vb u(t)
\end{equation}
with readout $y(t) = \vc^\top \phi(\vx(t))$. 
To approximate the constrained optimization problem, we use the loss
\begin{equation}
    \mathcal{L} = \alpha \mathbb{E}_z\left[\int_{\tau} ^{\tau + \xi} (y(t) - z)^2 dt\right] + \frac{1}{1+\lambda} \mathbb{E}_z\left[\int_0 ^T \|\phi(\vx(t))\|^2 dt\right] + \frac{\lambda}{1+\lambda} \left(\|\mW\|^2_F + \|\vb\|^2 \right),
\end{equation}
where we set the task term strength as $\alpha = 10^{3}$, and the weight and activity costs have been normalized by $1 + \lambda$ so that the weight of the constraint-enforcing terms relative to the task term is roughly consistent across $\lambda$. 

As in the linear case, we fix the readout weights to be unit norm, and simplify the optimization further by fixing it to be along the uniform direction as $\vc = \frac{1}{\sqrt{N}} \bm{1}$. We use networks of $N=80$ neurons and discretize the dynamics with a time constant timestep $dt = 0.05$. Networks were trained with the Adam optimizer at a learning rate of 1e-3 for $5 \times 10^5$ steps.

To compute the adjoint variable $\vq(t) = \frac{dy(\tau)}{d\vx(t)}$ for Fig. \ref{fig:4}e, we numerically integrated the backwards-in-time adjoint dynamics
\begin{align}
    -\frac{d \vq(t) }{dt} &= \mJ(\vx(t))^\top \vq(t)
\end{align}
with initial condition $\vq(\tau) =\text{diag}(\phi'(\vx(\tau)))\vc$, where $\mJ(\vx) = -\mI + \mW \text{diag}(\phi'(\vx))$ is the Jacobian of the forward dynamics. 
\subsection{ALM model and training}\label{sec:app_alm}
We train networks with $N=128$ neurons, and discretize the dynamics with timestep $dt = 0.25\uptau_x$ (units of time constants, with $\uptau_x$ = 40 ms).   
For the fixed reservoir used to preprocess the pulse inputs before passing as input to the trained networks, we use the same random reservoir parameters with $\tanh$ nonlinearity as in \citet{pereira-obilinovicNeuralDynamicsOutside2026} (described in https://github.com/ulisespereira/alm-coding-residuals-condavg-train), with dynamics
\begin{equation}
    \dot{\vz} = -\vz + \mW_{\text{octx}} \tanh(\vz) + \bm{I}_{\text{stim}}(t).
\end{equation}
The state of this reservoir is then passed as input $\vu(t) = \tanh(\vz(t))$ to the trained network through weights $\mB$.

We use the following specification for the timing of task periods:
\begin{itemize}
    \item $t_{\text{start}} = -2.1$
    \item sample cue onset at $t_{\text{sample}}=-1.85$ s consisting of three 150 ms pulses with 100 ms gaps in between, occurring over $[-1.85, -1.20]$ s
    \item delay period onset at -1.2 s
    \item go cue at 0 s
    \item Response window: $[t_{\text{resp}}=+0.1, +0.35]$ s
    \item heldout window: $[+0.1, t_{\text{end}}=0.6]$ s
\end{itemize}
We train networks with the objective
\begin{equation}
    \mathcal{L} = \alpha \mathcal{L}_{\text{task}} + \frac{1}{1+\lambda}\int_{t_{\text{sample}}}^{t_{\text{end}}} \|\phi(\vx)\|^2 dt + \frac{\lambda}{1+\lambda} \left(\|\mW\|_F^2 + \|\mB\|_F^2\right), 
\end{equation}
where $\mathcal{L}_{\text{task}} = \frac{\int_{t_{\text{start}}}^{t_{\text{resp}}} \sum_{c \in \{L, R\}} (y_c(t) - y^{\star}_c(t))^2 dt}{\int_{t_{\text{start}}}^{t_{\text{resp}}} \sum_{c \in \{L, R\}} y^{\star}_c(t)^2 dt}$. We choose a task term strength $\alpha = 10^{4}$, and use the Adam optimizer with learning rate lr=$2 \times 10^{-3}$ and cosine annealing to 0.05 lr over $8000$ epochs. We initialized recurrent weights and inputs weights as $\mW_{ij} \sim (10^{-3}/\sqrt{N} )\mathcal{N}(0,1)$, $\mB_{ia} \sim (1/\sqrt{200})\mathcal{N}(0,1)$, the biases as $\vb^0 = \bm{0}$, and the max firing rates as $r_i^{\text{max}} = 1$. For each combination of $\lambda \in \{0.1, 0.3, 1, 3, 10,30,100\}$ and $\sigma \in \{0, 10^{-3}, 10^{-2}\}$, we trained a total of $20$ networks with different random seeds. 
\subsubsection{Perturbation and analysis procedures}\label{sec:app_pert}

To estimate task coding directions, we averaged the condition difference direction $\vr_R(t) - \vr_L(t)$ over the following periods:
\begin{itemize}
    \item sample: first 0.6 s of the sample period
    \item choice: last 0.6 s of the delay period
    \item response: 0.6 s following the onset of the go cue
\end{itemize}
Perturbations were applied during the early delay period at a time $t_0 \in \{-1.1, -0.9, -0.7\}$ and in each trial condition as a shift to the state $\vx(t_0) \rightarrow  \vx(t_0) + g \hat{\vv}$, where $\hat{\vv}$ is a unit direction representing the perturbed direction with a signed perturbation strength $g \in \pm \{2,4,6,8,10,12\}$. To determine whether or not a flip of the choice occurred, we use a similar criterion as in \citep{pereira-obilinovicNeuralDynamicsOutside2026}: a trial with condition $c \in \{L, R\}$ was considered flipped by a perturbation if the network output averaged over the last $50$ timesteps $\overline{y}_c$ satisfied $|\overline{y}_c - \overline{y}_{\neg c}^\star| < \varepsilon$, where $\varepsilon = 0.25 |\overline{y}^\star_R -\overline{y}^\star_L|$. The flip rate was then measured as the fraction of all combinations of $g$, trial conditions, and perturbation times $t_0$ that resulted in a flip.   

For Fig. \ref{fig:5}b, we report nonnormality as Henrici's departure 
\begin{equation}
    \frac{\sqrt{\|\mW\|^2_F - \sum_{i}|\lambda_i(\mW)|^2}}{\|\mW\|_F},
\end{equation}
where $\lambda_i(\cdot)$ represents the $i$th eigenvalue. 
\clearpage
\section{Supplementary figures}
\begin{figure}[!thtb]
    \centering
    \includegraphics[width=0.8\linewidth]{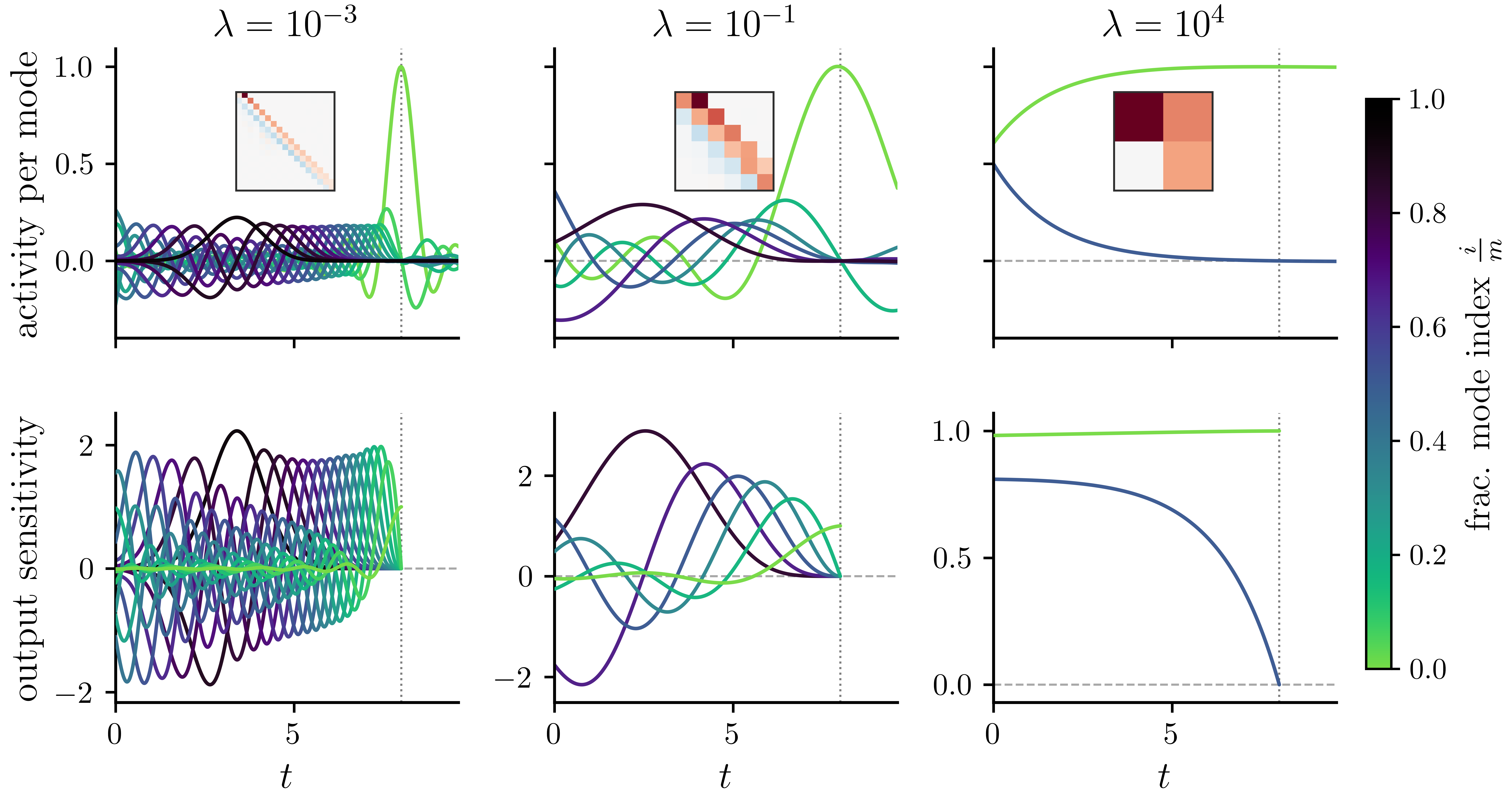}
    \caption{Analogous to Fig. \ref{fig:2}a, but with an adjoint cost $\gamma = 0.01$.}
    \label{fig:supp_modes}
\end{figure}

\begin{figure}[!htb]
    \centering
    \includegraphics[width=0.49\linewidth]{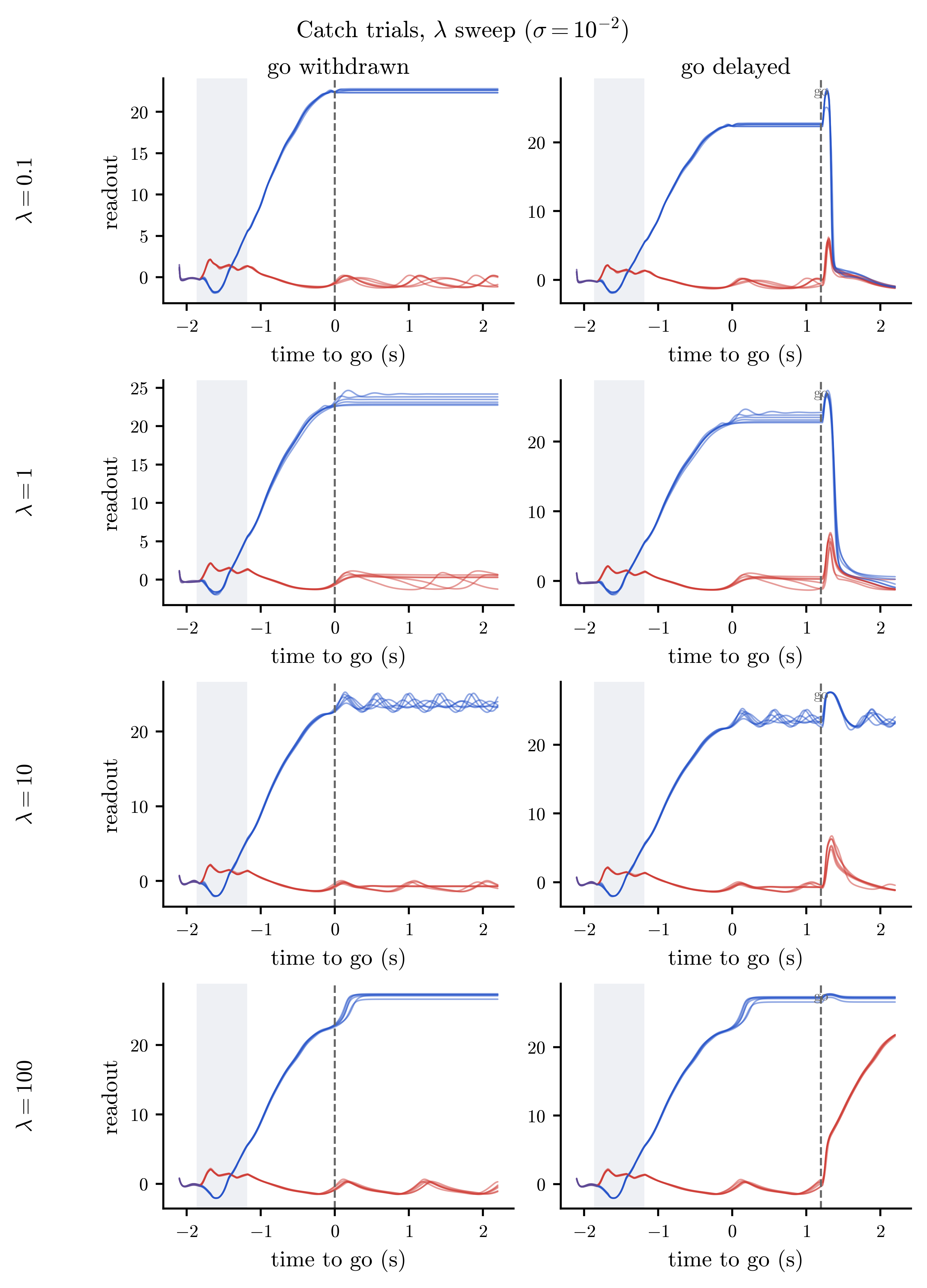}\hfill\includegraphics[width=0.49\linewidth]{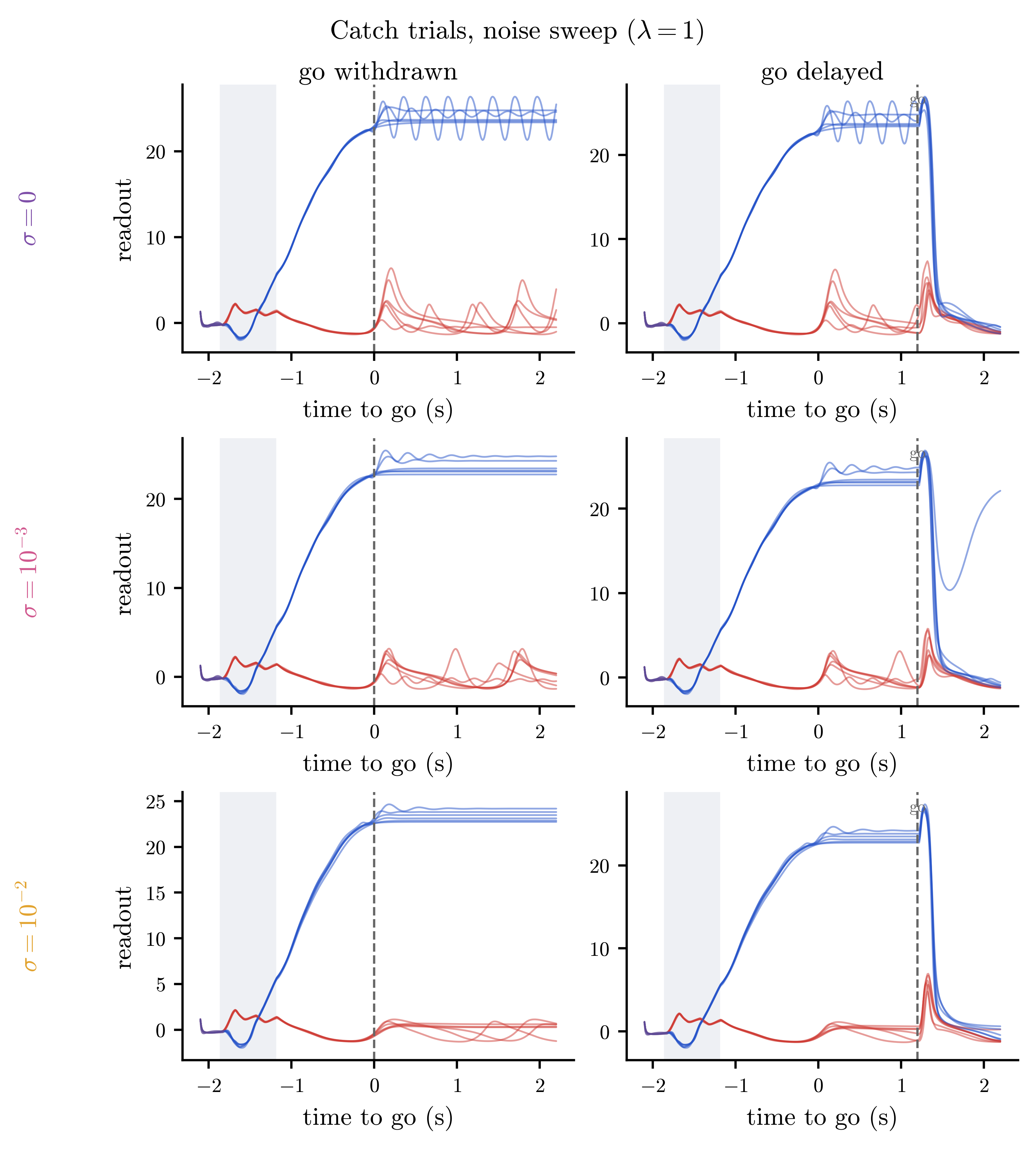}
    \caption{Catch trials where the go cue is either withdrawn entirely or delayed by 1.2 s, shown across $\lambda$ at fixed $\sigma = 10^{-2}$ (left) and across $\sigma$ at fixed $\lambda = 1$ (right).}
    \label{fig:supp_catch}
\end{figure}

\begin{figure}[!htb]
    \centering
    \includegraphics[width=\linewidth]{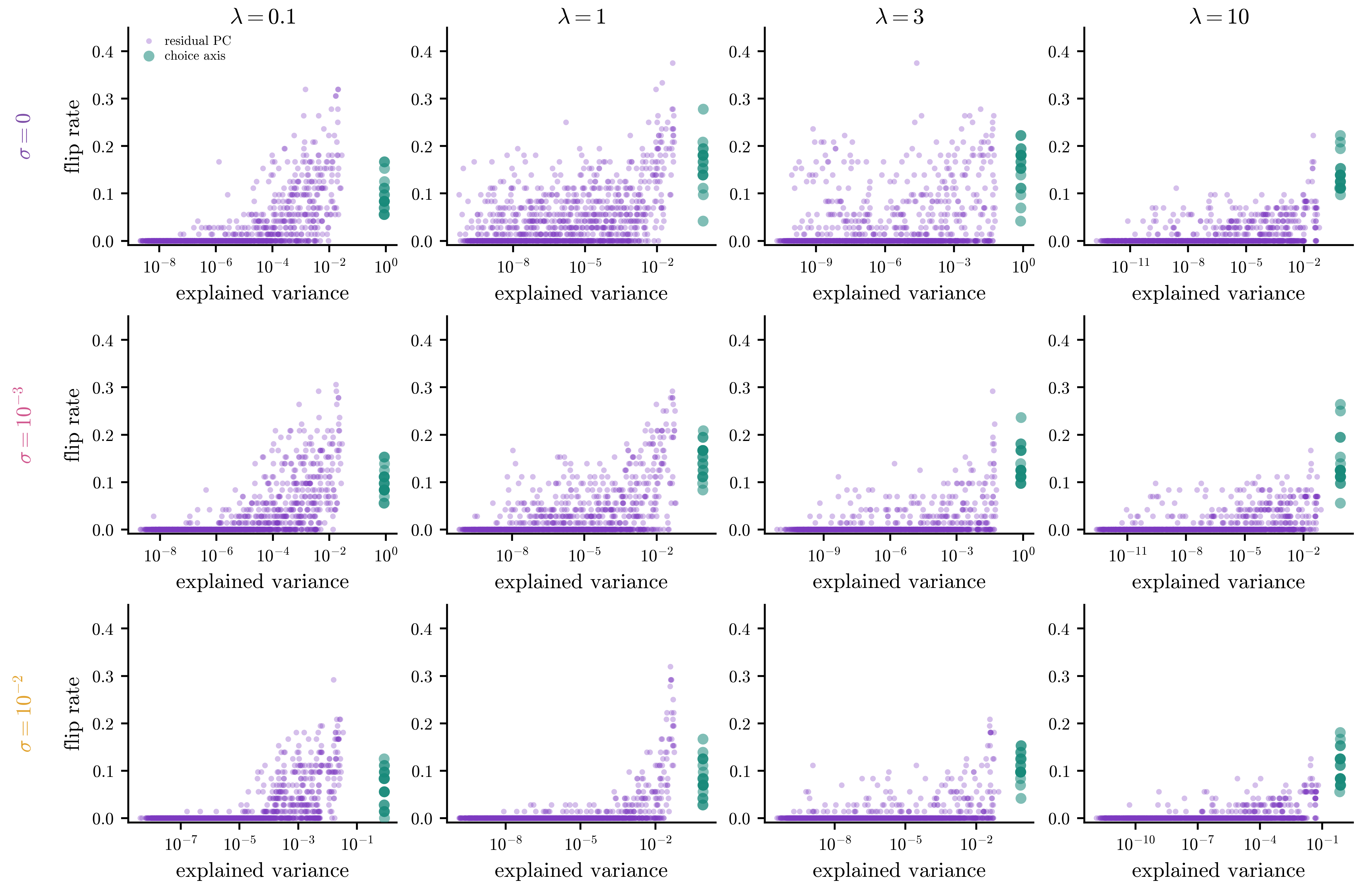}
    \caption{Sweep of explained variance versus flip rate across $\sigma$ and $\lambda$. }
    \label{fig:supp_evflip}
\end{figure}

\end{document}